\documentclass[aps,prl,twocolumn,superscriptaddress,showpacs,showkeys]{revtex4-1}

\usepackage{graphicx}
\usepackage{amsmath, amsxtra, amssymb, mathtools, isomath, nccmath, xspace, siunitx}
\usepackage{color}
\usepackage[utf8]{inputenc}
\usepackage[british]{babel}
\usepackage{hyperref}
\usepackage{sidecap}

\newcommand{\ket}[1]{\ensuremath{\lvert #1 \rangle}\xspace}%
\newcommand{\bra}[1]{\ensuremath{\langle #1 \rvert}\xspace}%
\newcommand{\braket}[2]{\ensuremath{\langle #1 \rvert #2\rangle}\xspace}%
\newcommand{\expect}[3]{\ensuremath{\langle #1 \lvert #2 \rvert #3 \rangle}\xspace}%
\newcommand{\alat}{\ensuremath{a_{\text{lat}}}\xspace}%
\sidecaptionvpos{figure}{h}

\begin{document}
\title{\bf{Coherent many-body spin dynamics in a long-range interacting Ising chain}} 

\author{Johannes Zeiher}
\email[]{johannes.zeiher@mpq.mpg.de}
\affiliation{Max-Planck-Institut f\"{u}r Quantenoptik, 85748 Garching, Germany}

\author{Jae-yoon Choi}
\affiliation{Max-Planck-Institut f\"{u}r Quantenoptik, 85748 Garching, Germany}

\author{Antonio Rubio-Abadal}
\affiliation{Max-Planck-Institut f\"{u}r Quantenoptik, 85748 Garching, Germany}

\author{Thomas Pohl} 
\affiliation{Department of Physics and Astronomy, Aarhus University, DK 8000 Aarhus C, Denmark}

\author{Rick van Bijnen}%
\affiliation{Institut f\"ur Quantenoptik und Quanteninformation,\"Osterreichische Akademie der Wissenschaften, 6020 Innsbruck, Austria}

\author{Immanuel Bloch}%
\affiliation{Max-Planck-Institut f\"{u}r Quantenoptik, 85748 Garching, Germany}
\affiliation{Fakult\"{a}t f\"{u}r Physik, Ludwig-Maximilians-Universit\"{a}t M\"{u}nchen, 80799 M\"{u}nchen, Germany}%

\author{Christian Gross}%
\affiliation{Max-Planck-Institut f\"{u}r Quantenoptik, 85748 Garching, Germany}

\date{\today}


\begin{abstract}
Coherent many-body quantum dynamics lies at the heart of quantum simulation and quantum computation.
Both require coherent evolution in the exponentially large Hilbert space of an interacting many-body system~\cite{Lloyd1996,Lechner2015}.
To date, trapped ions have defined the state of the art in terms of achievable coherence times in interacting spin chains~\cite{Blatt2012b,Jurcevic2014,Richerme2014,Bohnet2016a}.
Here, we establish an alternative platform by reporting on the observation of coherent, fully interaction-driven quantum revivals of the magnetization in Rydberg-dressed Ising spin chains of atoms trapped in an optical lattice. We identify partial many-body revivals at up to about ten times the characteristic time scale set by the interactions.
At the same time, single-site-resolved correlation measurements link the magnetization dynamics with inter-spin correlations appearing at different distances during the evolution. These results mark an enabling step towards the implementation of Rydberg atom based quantum annealers~\cite{Glaetzle2016}, quantum simulations of higher dimensional complex magnetic Hamiltonians~\cite{Glaetzle2015,vanBijnen2015}, and itinerant long-range interacting quantum matter~\cite{Henkel2010,Mattioli2013,Geissler2015}.
\\
\end{abstract}

\maketitle


The coherent unitary evolution of closed many-body quantum systems initially prepared in
a superposition of different eigenstates is one of the most fundamental concepts
of quantum theory. It predicts a fast dephasing of the initial state,
followed by its revival after long times. This dynamics originates from
the discrete energy spectrum of the many-body eigenstates, each evolving with
its characteristic frequency. The expected collapse and revival dynamics is in
stark contrast to the experience, that in typical interacting macroscopic systems quantum revivals
are entirely absent. This is a consequence of the exponential increase of
the number of distinct energy levels with system size, making the spectrum effectively continuous,
such that the revival time diverges. Even in small generic interacting
systems with few constituents, the observation of quantum revivals is far from
trivial due to residual couplings to the (macroscopic) environment causing decoherence.
This renders the observation of quantum revivals in a many-body system one of the most
demanding tests to demonstrate its coherent evolution, which itself is indispensable, for
example, for efficient adiabatic quantum computation~\cite{Lloyd1996,Lechner2015}.

Quantum revivals have so far been observed experimentally using appropriate
observables in few-particle systems, whose dynamics was additionally constrained
to small parts of the Hilbert space.
Seminal achievements are the observation of collapse and revivals of Rabi oscillation dynamics of
Rydberg states coupled to the radiation field in a microwave cavity~\cite{Rempe1987, Brune1996},
of revivals of the motional state of a single trapped ion~\cite{Meekhof1996} or of coherent
photonic states in microwave resonators coupled to a
transmon qubit ~\cite{Kirchmair2013}. Revival dynamics has also been observed for small
coherent states of matter realized with ultracold atoms in individual sites of
optical lattices. The onsite interactions between the indistinguishable atoms
led first to dephasing and later to revivals of the relative phase of the
condensate wave function on the different lattice sites~\cite{Greiner2002,Will2010}.

A much more complex many-body dynamics is expected when the full Hilbert space
is accessible in the time evolution, and in particular in systems with long-range
interactions. Quantum magnets featuring such interactions have been realized
recently with trapped ions~\cite{Blatt2012b,Jurcevic2014,Richerme2014,Bohnet2016a}, ground state
molecules~\cite{Yan2013}, magnetic atoms~\cite{dePaz2013a} and neutral atoms
coupled to Rydberg states on resonance~\cite{Schauss2015,Labuhn2016a}, or 
off-resonantly~\cite{Jau2016,Zeiher2016a}. Indeed, indications for partial quantum
revivals have been detected in the dynamics of the next-nearest-neighbor
correlations~\cite{Richerme2014} and in the magnetization of small systems of
three spins~\cite{Barredo2015a}.

Here we report on the observation of quantum revivals of the transverse
magnetization in Rydberg-dressed Ising chains of about ten atoms with long-range interactions.
The magnetization of the chains shows partial revivals
up to times exceeding ten times the characteristic timescale set by the nearest-neighbor
coupling strength and single-spin resolved correlation measurements reveal the
microscopic origin of the interaction-induced collapse dynamics at short times.

\begin{figure*}
  \centering
  \includegraphics{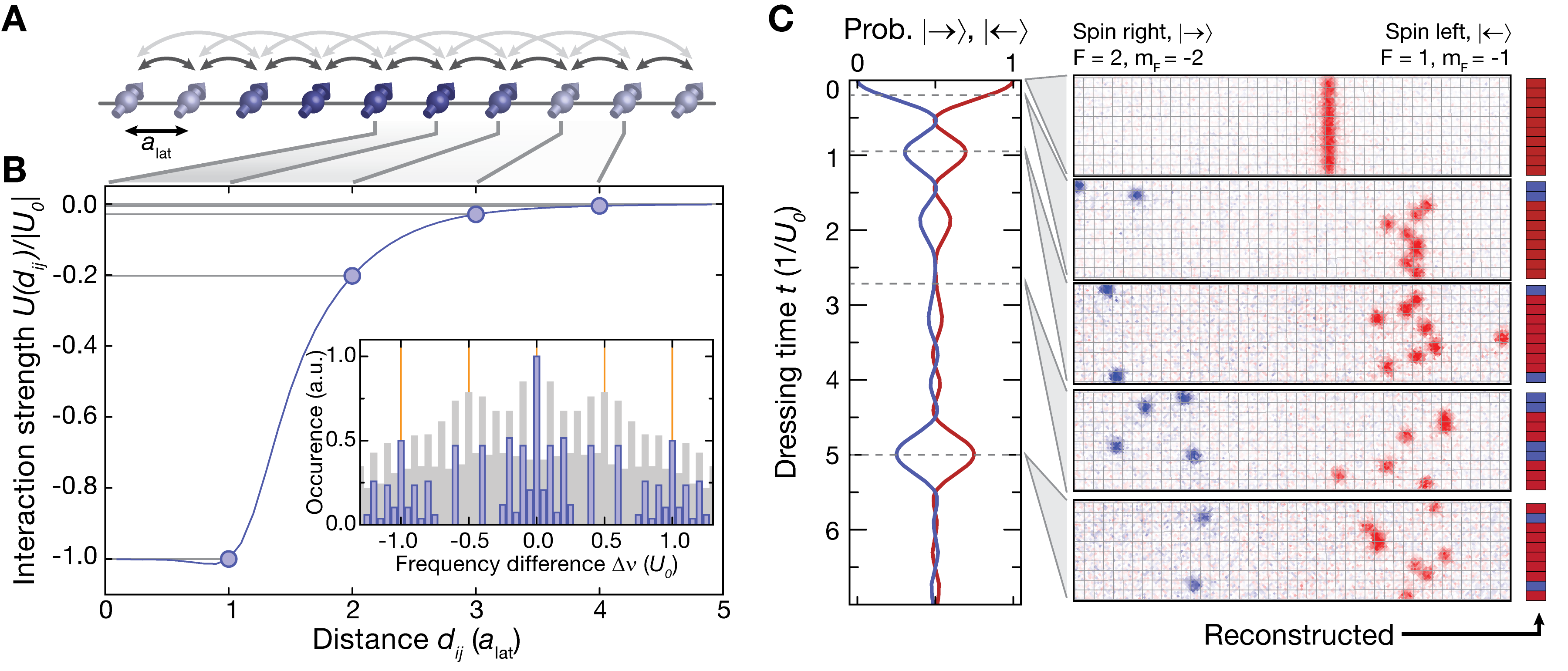}%
  \caption{ \label{fig:1}
  \textbf{Schematic of the Ising spin chain and spin-selective detection.}
  \textbf{(A)} Illustration of a spin chain with $N=10$ spins initialized in the
  fully transverse magnetized state. The dominant contributions of the
  Rydberg-dressed interaction between nearest and next-nearest spins spaced by
  $\alat$ is indicated by the dark gray and light gray arrows. The brightness of the blue color of the
  spins encodes the interaction strength between the exemplary selected fifth
  spin with the rest of the chain. The gray lines are guides to the eye to link
  to the interaction potential shown in B. \textbf{(B)} Calculated
  Rydberg-dressed potential normalized to the nearest-neighbor interaction strength
  $|U_0|=13.1(5)\,$kHz (blue solid line) with the relevant potential at
  multiples of the lattice distance $\alat$ marked by blue points and gray
  horizontal lines. The inset shows the occurrence of frequency differences 
  $\Delta\nu$ in the many-body spectrum of the long-range interacting Ising
  model (gray bars) and those governing transverse magnetization dynamics (blue bars).
  Orange vertical lines mark the corresponding $\Delta\nu$ for the 
  Ising model with nearest-neighbor interactions only. \textbf{(C)} Simulated revival dynamics of
  the populations of spin left (red) and spin right (blue) starting
  from the initially fully magnetized chain in $S^y$-direction for a defect-free chain
  of $10$ spins and long-range interactions. Clear partial revivals are observed during the evolution. The
  fluorescence images to the right show characteristic spin configurations for
  $10$ spins observed during the collapse and revival dynamics at times
  indicated by the gray lines. The spin of the atoms was detected via an in-situ
  Stern-Gerlach sequence, which lead to a spatial separation of spin left
  (red) and spin right atoms (blue). This allows for the reconstruction of the
  full spin and density distribution (pictograms to the right).}
\end{figure*}

In our experiments, we implemented a one-dimensional (1d) Ising spin chain
of $N$ spins with soft-core type long-range interactions using ultracold rubidium-$87$
atoms in an optical lattice, optically ``dressed'' to a strongly interacting Rydberg
state~\cite{Santos2000,Bouchoule2002,Henkel2010}, see Fig.~\ref{fig:1}A. The spin-1/2
degree of freedom is encoded in two hyperfine ground states
and its dynamics is described by the Hamiltonian
\begin{equation}
\hat{H}=h\sum_{i\neq j}^{N}\frac{U(d_{ij})}{2}~\hat{S}^z_i \hat{S}^z_j.
\end{equation}
Here, $\hat{S}^{z}_i$ denotes the spin operator measuring the spin in the
$z$-direction at a lattice site $i$ and we have omitted all terms linear in the spin operators
as they are irrelevant for the subsequent discussion~\cite{SI}. For the chosen parameters
of the optical coupling, the interaction potential $U(d_{ij})$ is approximated by a soft-core
shape for spins at a distance $d_{ij}=|i-j|$ (see Fig.~\ref{fig:1}B). For distances
smaller than the lattice spacing $d_{ij}\leq a_\mathrm{lat}=532\,$nm
it saturates to the nearest-neighbor value $U(1)=U_0\approx-13\,$kHz and for larger
distances it asymptotically falls off with a van-der-Waals tail, $U(d_{ij})\propto1/d_{ij}^6$
~\cite{Zeiher2016a}.

To study the quantum evolution, the system is initially prepared in a separable coherent
spin state~\cite{Radcliffe1971} with maximal magnetization 
along the $S^y$-direction,
$\ket{\psi_0}=(\ket{\leftarrow})^{\otimes N}$, where $\ket{\leftarrow}$ is a
single-spin eigenstate of $\hat{S}^y$. In the $\hat{S}^z$ basis, in
which all many-body eigenstates of $\hat{H}$ factorize, each
spin is equally likely found in the two single-spin eigenstates of $\hat{S}^z$. Thus, each
many-body eigenstate $\ket{\lambda}$ with possibly degenerate eigenenergy
$E_\lambda=h\nu_\lambda$ is populated with equal probability $\lvert\braket{\lambda}{\psi_0}\rvert^2=|c_\lambda|^2$, and an amplitude of
$c_\lambda=2^{-N/2} e^{i\phi_\lambda}$. After unitary evolution of $\ket{\psi_0}$ with $\hat{U}=e^{-i\hat{H}t/\hbar}$, the expectation value of the local transverse magnetization at site $j$ in the chain becomes $\langle\hat{S}^y_j(t)\rangle=2^{-N}\sum\limits_{\lambda,\eta}
e^{-i 2\pi\left( \nu_{\lambda}-\nu_{\eta}\right) t} \expect{\eta}{\hat{S}^y_j}{\lambda}$, where the irrelevant phases $\phi_{\lambda}$ and $\phi_{\eta}$ have been omitted. The dynamics
of $\langle\hat{S}^y_j(t)\rangle$ is hence determined by those frequency differences $\Delta\nu=\nu_\lambda-\nu_\eta$, for which the operator matrix-element $\expect{\eta}{\hat{S}^y_j}{\lambda}$ does not vanish (see Fig.~\ref{fig:1}B).
Due to the contribution of different $\Delta\nu$, the initial state and hence the magnetization is expected to undergo an initial collapse dynamics, followed by a revival on longer timescales. Importantly, this dynamics is purely driven by the build-up of correlations~\cite{Bohnet2016a}, since the initial state is a mean-field steady state with $\langle\hat{S}^z_j\rangle=0$.
Indeed, the collapse is accompanied by the generation of entanglement, at short
evolution times in the form of spin squeezing~\cite{Hazzard2014,Macri2016a,Foss-Feig2016}.
At a later time $t=1/2U_0$, the system realizes approximately a cluster
state~\cite{Raussendorf2001}, a highly entangled state useful for quantum
computation.

\begin{figure*}
  \centering
  \includegraphics{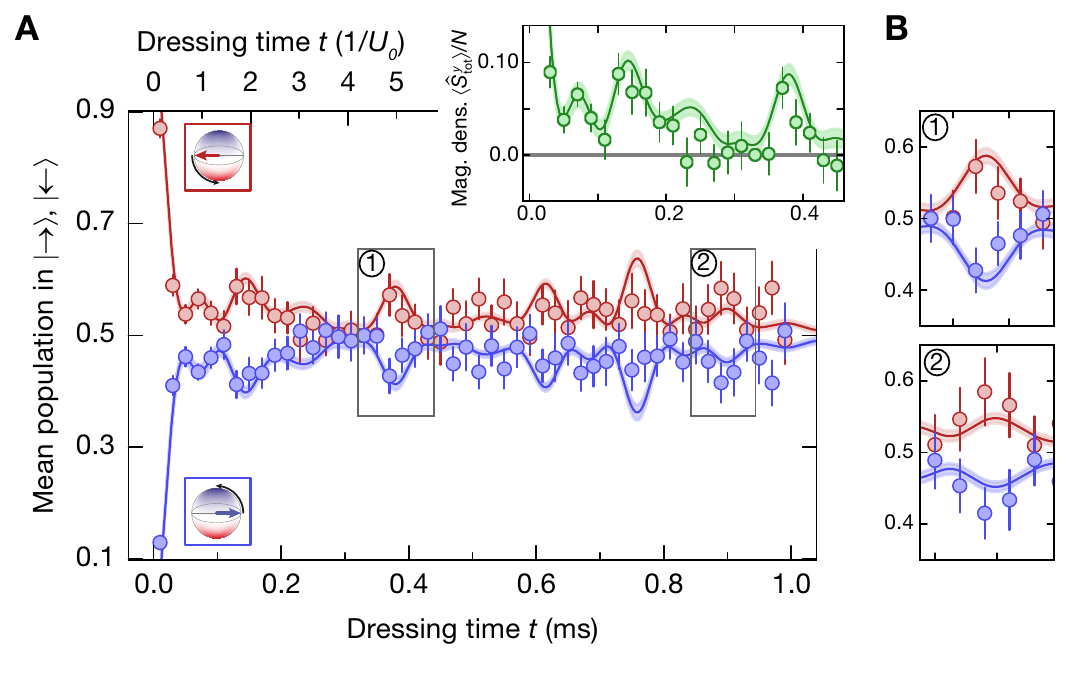}%
  \caption{ \label{fig:2}
  \textbf{Evolution of the mean magnetization density.} \textbf{(A)} The probability
  to measure atoms in the states $\ket{\leftarrow}$ ($\ket{\rightarrow}$)
  versus total dressing time $t$ is shown as red (blue) data points. The final spin rotation before detection is indicated
  by the pictograms (upper and lower left corner). The upper axis is scaled in units of the inverse
  nearest-neighbor interaction $1/|U_0|=76(3)\,\mu$s. The total atom number was restricted to be
  $N<15$ to filter out events with clear preparation errors. The solid line
  shows the theoretically expected dynamics, averaged over $100$ initial chains randomly
  selected from a reference dataset with an initial filling of $87(3)\%$ and a mean atom
  number of $10(1.4)$. The shaded region marks the corresponding
  standard error of the mean (s.e.m.). All data points are an average over at
  least $50$ experimental realizations ($150$ for $t<0.1\,$ms). The inset shows the initial dynamics of the
  mean transverse magnetization density $\langle\hat{S}^y_{\mathrm{tot}}\rangle/N$ up to $t\approx6/U_0$ obtained
  from the spin populations (green data points) with the theoretical prediction (green solid line).
  \textbf{(B)} Zoom into two revival features at $t\approx5/U_0$ (1) and $t\approx12/U_0$ (2), as indicated by the
  gray boxes in (A). 
  All error bars on the data points denote one s.e.m..} 
\end{figure*}

To obtain an intuitive understanding of the collapse and revival dynamics,
it is illustrative to consider the case of an Ising model with only nearest-neighbor interactions of
strength $h U_0$. For such a case, the transverse magnetization $\langle\hat{S}^y_{\mathrm{tot}}\rangle=\sum\limits_{{j=1}}^N \langle\hat{S}^y_j\rangle$ would show periodic revivals at times $t_{n}=n/U_0$. This is due to the highly degenerate spectrum with all relevant energy differences $\Delta\nu$ either vanishing, or being equal to $U_0$ for a spin located in the bulk or to $U_0/2$ for a spin at the edge of the system (see Fig.~\ref{fig:1}B and ~\cite{SI}).
Adding interactions with longer range leads to a more complex spectrum, breaking many of the degeneracies present in the former case. For the interaction potential realized in the experiment and a chain of $N=10$ atoms,
also the number of relevant frequency differences increases significantly (see Fig.~\ref{fig:1}B). Whereas this shifts a possible perfect revival of the initial state to experimentally inaccessible times, the magnetization may still show dynamics with partial revivals (see Fig.~\ref{fig:1}C). The exact magnetization dynamics can be calculated analytically for our initial state and it predicts that the local transverse magnetization evolves in time as $\langle\hat{S}^y_j(t)\rangle=\frac{1}{2}\prod\limits_{i\neq
j}^{N}\cos(\pi U(d_{ij})t)$~\cite{Foss-Feig2013,Hazzard2014}. This confirms the
intuition that partial revivals of the magnetization of a single spin originate from its interaction $U(d_{ij})$ with spins at different distances $d_{ij}$ and the resulting beat notes.
Despite the conceptual similarities to previously studied revival
dynamics~\cite{Brune1996,Meekhof1996,Greiner2002,Will2010,Kirchmair2013}, an
important difference of the spin system with long-range interactions studied here is the
absence of a spatial spin exchange symmetry and, hence, there is no simplifying description of the system's
temporal dynamics in terms of symmetric Dicke states~\cite{Schachenmayer2013}. Thus, here the 
dynamics is in a regime exploring a much larger portion of the Hilbert space, leading also to an interesting spatial structure in the time evolution, that we reveal by our microscopic detection (see
Fig.~\ref{fig:1}C and ~\cite{SI}).


Our experiments started with the preparation of an atomic chain of ten sites
with $87(3)\%$ filling from a Mott insulator using single-site addressing
techniques~\cite{Weitenberg2011a,Fukuhara2013,SI}.  Subsequently, the coherent spin state
was initialized by a global microwave-induced $\pi/2$ rotation about the
$S^x$-axis ending in an equal superposition of the hyperfine states
$\ket{F,~m_F}=\ket{1,-1}$ and $\ket{2,-2}$. Next, we switched on the interactions
$U(d_{ij})$ for a total time $t$ by illuminating the sample with the ``dressing''
laser, coupling to the $31P_{1/2}$ Rydberg state with Rabi frequency
$\Omega/2\pi=3.57(3)\,$MHz and detuning $\Delta/2\pi=11.00(2)\,$MHz. This
Rydberg dressing was interrupted after a time $t/2$ to implement a spin-echo
pulse of area $\pi$ about the $S^x$-axis to remove trivial phases accumulated due
to single atom shifts proportional to $\hat{S}^z$, thereby leaving the spin-spin interaction as the only drive of the dynamics~\cite{SI}. Finally, the spin along the $S^y$-direction was read out using a
second $\pi/2$-pulse to rotate the spins to the $\hat{S}^z$-basis, where we separated
the spins in-situ making use of their different magnetic
moments. This enabled a position-dependent readout of the spin direction~\cite{Fukuhara2015} and provided access to the local and global magnetization as well as the total atom number $N$ after the dynamics.

\begin{figure}
  \centering
  \includegraphics{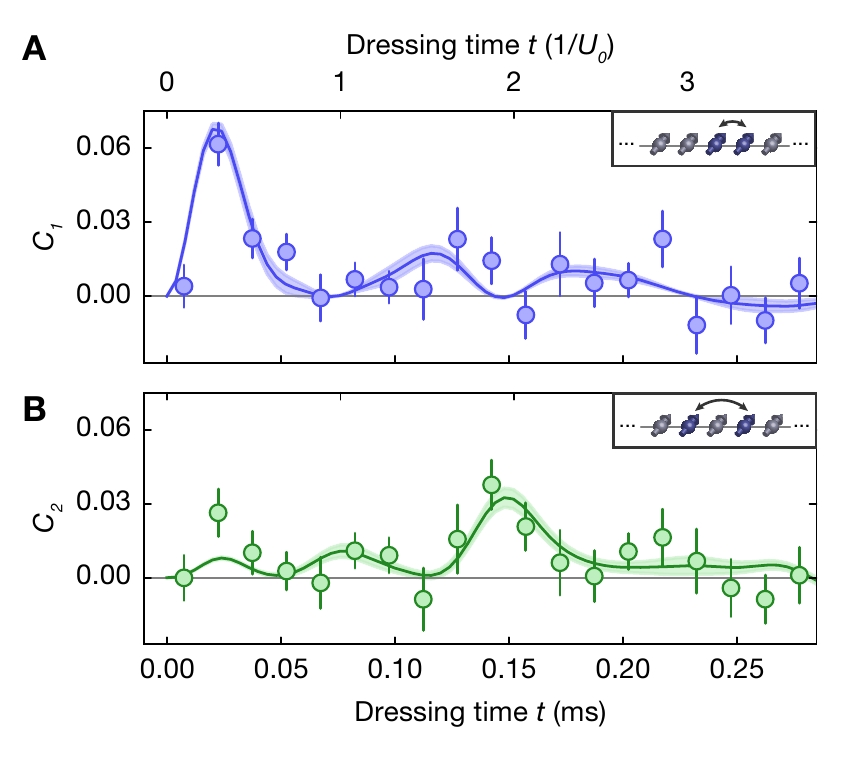}%
  \caption{\label{fig:3}
  \textbf{Local characterization of the spin dynamics.} \textbf{(A)} Observed evolution of the
  connected correlations $C_d$ for spins separated by $d=\alat$ and \textbf{(B)}
  $d=2\alat$, as indicated by the pictograms. The solid line with surrounding shading
  shows the corresponding prediction of the same theoretical calculation as
  used in Fig.~\ref{fig:2}. All error bars denote one s.e.m..}
\end{figure}

The collapse and revival dynamics of the initial coherent spin state is observed by
tracking the evolution of the spin populations along the $S^y$-direction. This is equivalent to tracking the
mean transverse magnetization density $\langle\hat{S}^{y}_{\mathrm{tot}}(t)\rangle/N$, where the
measured total atom number at time $t$ is used for the normalization (see Fig.~\ref{fig:2}).
For short times, we find a fast interaction-induced depolarization dynamics up to a time of approximately $t\approx
40\,\mu\mathrm{s} \approx 1/2U_0$, before exhibiting clear partial
revivals at $t\approx1/U_0$, $2/U_0$ and $5/U_0$. The times of the revivals can
be qualitatively explained by considering only the nearest and next-nearest
neighbor interaction, which differ by a factor of about five in our case. Interestingly, even at longer
times, the data closely follows the numerical prediction showing a finite
magnetization signal up to $t=12/U_0$, approaching the $1/e$-lifetime
$\tau=1.21(3)\,$ms of the atom number in the chains. This indicates that the atom loss does not induce excessive dephasing, which would result in a vanishing magnetization density in the system. 
Our numerical simulation of the dynamics considers pure unitary evolution and additionally takes into account the randomness due to the initially imperfect filling of the chain. The presence of defects in the chain suppresses all revivals at odd multiples of $1/U_0$ as an empty site causes
the neighboring spin to evolve with approximately half the frequency~\cite{SI}.

Investigating spatial spin-spin correlations provides microscopic insight into
the collapse and revival dynamics. To this end, we evaluated the connected
spin-spin correlator
$C_d=\langle\hat{S}^y_i\hat{S}^y_{i+d}\rangle-\langle\hat{S}^y_i\rangle\langle\hat{S}^y_{i+d}\rangle$
at different distances $d$ versus time
(see Fig.~\ref{fig:3}). The correlation signal for neighboring spins ($C_1$)
reveals that the initial decay of the magnetization is driven by the build-up of
strong nearest-neighbor correlations, peaking at $t\approx1/4U_0$.
Whereas for pure nearest-neighbor interactions one solely expects correlations
extending to the neighboring spins and $C_2=0$ for all times, we also find a
non-zero next-nearest-neighbor correlator $C_2$, peaking later at $t\approx 2/U_0$.
In our case, a non-zero $C_2$ is therefore a direct manifestation of the
long-range interactions. These can entangle distant parts of the system even
in the absence of moving quasi-particles~\cite{Cheneau2012,Richerme2014,Jurcevic2014},
which arise for example in the Ising model with transverse field 
as free fermions after a Jordan-Wigner transformation~\cite{Schachenmayer2013}.
An illustrative interpretation of the initial strong growth of the nearest-neighbor
correlation can be obtained by rewriting $\hat{H}_{\mathrm{int}}$ in
terms of the raising and lowering operators with respect to the $S^y$-direction,
$\hat{\tilde S}^{\pm}_j=\hat{S}^x_j\pm i\hat{S^z_j}$, yielding
$\hat{H}=h\sum\limits_{i\neq j}^{N} \frac{U(d_{ij})}{4}\left(\hat{\tilde
S}^+_i\hat{\tilde S}^-_j + \hat{\tilde S}^-_i\hat{\tilde S}^+_j -\hat{\tilde
S}^+_i\hat{\tilde S}^+_j -\hat{\tilde S}^-_i\hat{\tilde S}^-_j\right)$.
Starting with $\ket{\psi_0}$, where all spins are in state $\ket{\leftarrow}$,
only the last term in the sum contributes for short times, which flips spins
pairwise. This ``pair production'' of $\ket{\rightarrow}$ spins is strongest at
short distances, explaining the observed correlation signal qualitatively.

\begin{figure}
  \centering
  \includegraphics{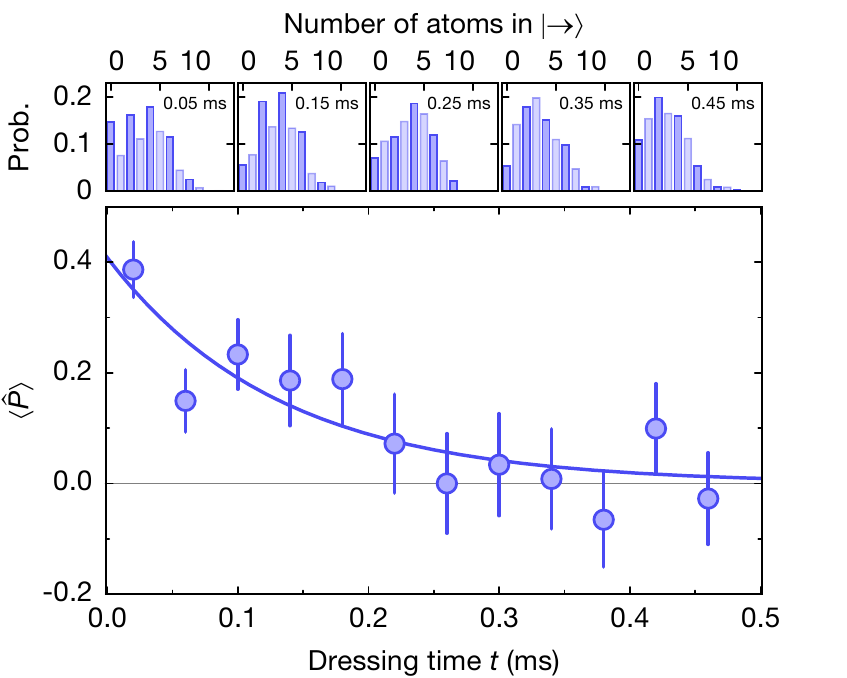}%
  \caption{\label{fig:4}
  \textbf{Evolution of the spin parity in the initially unpopulated $\ket{\rightarrow}$ state.}
  The histograms on top show the evolution of the number of detected atoms in
  the $\ket{\rightarrow}$ state with even numbers highlighted by the darker
  color. The data was binned in intervals of $0.1\,$ms and the bin centers are
  indicated. The main plot shows the extracted parity $\langle \hat{P} \rangle$ versus
  time together with an exponential fit with time constant $\tau_P=0.13(4)\,$ms (solid line).
  All error bars denote one s.e.m..}
\end{figure}

All terms in the Hamiltonian above are parity conserving since the number of
spins flipped from $\ket{\leftarrow}$ to $\ket{\rightarrow}$ can only be changed
in steps of two. Hence, ideally we expect to observe only even numbers of
$\ket{\rightarrow}$ spins to appear during the evolution. This situation is
similar to spontaneous pair creation in the squeezed vacuum state of photons.
There, a strongly parity-modulated signal in the excitation
numbers~\cite{Breitenbach1997} has also been observed, and interpreted as
quantum interference in phase space~\cite{Schleich1987}. The parity signal is
directly visible in the histograms shown in Fig.~\ref{fig:4} for different
evolution times and most pronounced at short times, where spurious single-spin
rotation terms proportional to $\hat{S}^z_i$ are smallest and atom number decay
is negligible. We characterized the evolution of the parity of the number of
detected atoms in $\ket{\rightarrow}$ by evaluating $\langle \hat{P} \rangle=\langle e^{-i\pi\sum
\limits_{i=1}^{N}{\hat{S}^{\rightarrow}_i}} \rangle$, with
$\hat{S}^{\rightarrow}_i=\ket{\rightarrow}_i\bra{\rightarrow}_i$ detecting if
atom $i$ is in the $\ket{\rightarrow}$ state. For the initial state at $t=0$,
we obtain $\langle \hat{P} \rangle=0.41(7)$, where the reduction of the parity
compared to unity is expected due to imperfections in the microwave rotations
and decoherence due to magnetic field noise, both leading to uncorrelated rotations
of individual spins even in the absence of Rydberg dressing~\cite{SI}. Subsequently,
the observed parity signal decays with increasing dressing time $t$ and we extract
a time constant of $\tau_P=0.13(4)\,$ms, which is on the order of $\tau/N$,
the characteristic time to lose a single atom. Hence, we conclude that off-resonant
excitation to the Rydberg state followed by a loss of the excited atom is the
dominating decoherence effect~\cite{SI}.
    
Our observations of coherent, interaction-driven collapse and revival dynamics
establish Rydberg dressing as a promising technique to study interacting spin
systems also in more complex scenarios~\cite{Glaetzle2015,vanBijnen2015}.
Indeed, the demonstrated interaction to decay ratio of $2\pi U_0
\tau\approx100$ exceeds our previous result~\cite{Zeiher2016a} by a factor of
$100$ and is comparable to the state of the art for implementing spin
Hamiltonians in ion chains~\cite{Jurcevic2014,Richerme2014,Zhang2017}.
While in the current experiment faster dynamics at larger interaction strengths
was not accessible due to merely technical limitations in the pulse timing accuracy,
the interaction strength and hence the interaction to decay ratio can be significantly
increased by working closer to resonance at larger Rydberg state admixture.
The improved available long coherence times combined with the controlled,
optically induced interactions allow for the implementation of Loschmidt-echo
type sequences to characterize the value of the generated states for quantum
metrological applications~\cite{Macri2016a}. The study of periodically driven
systems in two dimensions or in systems with periodic boundary conditions are
now also within reach, holding promise to shed new light on many-body phases existing
solely in non-equilibrium scenarios~\cite{Khemani2016b,Potirniche2016a}, among them the ``Floquet time crystal''
phase~\cite{Else2016c,Zhang2017,Choi2017}. Furthermore, our experiments mark
the first step towards a Rydberg quantum annealer based on
coherent Rydberg-dressed interactions~\cite{Glaetzle2016} and the study of itinerant quantum
matter with long-range soft-core interactions~\cite{Henkel2010,Mattioli2013,Geissler2015}.

\begin{acknowledgements}
We thank Simon Hollerith for proofreading the manuscript. We acknowledge support by the DNRF through a Niels Bohr Professorship for T.P. as well as funding by MPG, EU (UQUAM, RYSQ, Marie Curie Fellowship to J.C.), DFG through the SPP 1929 (GiRyd) and the K\"orber Foundation. \\
\end{acknowledgements}

\bibliography{RydbergRevivals_revtex}

\begin{thebibliography}{49}%
\makeatletter
\providecommand \@ifxundefined [1]{%
 \@ifx{#1\undefined}
}%
\providecommand \@ifnum [1]{%
 \ifnum #1\expandafter \@firstoftwo
 \else \expandafter \@secondoftwo
 \fi
}%
\providecommand \@ifx [1]{%
 \ifx #1\expandafter \@firstoftwo
 \else \expandafter \@secondoftwo
 \fi
}%
\providecommand \natexlab [1]{#1}%
\providecommand \enquote  [1]{``#1''}%
\providecommand \bibnamefont  [1]{#1}%
\providecommand \bibfnamefont [1]{#1}%
\providecommand \citenamefont [1]{#1}%
\providecommand \href@noop [0]{\@secondoftwo}%
\providecommand \href [0]{\begingroup \@sanitize@url \@href}%
\providecommand \@href[1]{\@@startlink{#1}\@@href}%
\providecommand \@@href[1]{\endgroup#1\@@endlink}%
\providecommand \@sanitize@url [0]{\catcode `\\12\catcode `\$12\catcode
  `\&12\catcode `\#12\catcode `\^12\catcode `\_12\catcode `\%12\relax}%
\providecommand \@@startlink[1]{}%
\providecommand \@@endlink[0]{}%
\providecommand \url  [0]{\begingroup\@sanitize@url \@url }%
\providecommand \@url [1]{\endgroup\@href {#1}{\urlprefix }}%
\providecommand \urlprefix  [0]{URL }%
\providecommand \Eprint [0]{\href }%
\providecommand \doibase [0]{http://dx.doi.org/}%
\providecommand \selectlanguage [0]{\@gobble}%
\providecommand \bibinfo  [0]{\@secondoftwo}%
\providecommand \bibfield  [0]{\@secondoftwo}%
\providecommand \translation [1]{[#1]}%
\providecommand \BibitemOpen [0]{}%
\providecommand \bibitemStop [0]{}%
\providecommand \bibitemNoStop [0]{.\EOS\space}%
\providecommand \EOS [0]{\spacefactor3000\relax}%
\providecommand \BibitemShut  [1]{\csname bibitem#1\endcsname}%
\let\auto@bib@innerbib\@empty
\bibitem [{\citenamefont {Lloyd}(1996)}]{Lloyd1996}%
  \BibitemOpen
  \bibfield  {author} {\bibinfo {author} {\bibfnamefont {S.}~\bibnamefont
  {Lloyd}},\ }\href {\doibase 10.1126/science.273.5278.1073} {\bibfield
  {journal} {\bibinfo  {journal} {Science}\ }\textbf {\bibinfo {volume}
  {273}},\ \bibinfo {pages} {1073} (\bibinfo {year} {1996})}\BibitemShut
  {NoStop}%
\bibitem [{\citenamefont {Lechner}\ \emph {et~al.}(2015)\citenamefont
  {Lechner}, \citenamefont {Hauke},\ and\ \citenamefont
  {Zoller}}]{Lechner2015}%
  \BibitemOpen
  \bibfield  {author} {\bibinfo {author} {\bibfnamefont {W.}~\bibnamefont
  {Lechner}}, \bibinfo {author} {\bibfnamefont {P.}~\bibnamefont {Hauke}}, \
  and\ \bibinfo {author} {\bibfnamefont {P.}~\bibnamefont {Zoller}},\ }\href
  {\doibase 10.1126/sciadv.1500838} {\bibfield  {journal} {\bibinfo  {journal}
  {Science Advances}\ }\textbf {\bibinfo {volume} {1}},\ \bibinfo {pages}
  {e1500838} (\bibinfo {year} {2015})}\BibitemShut {NoStop}%
\bibitem [{\citenamefont {Blatt}\ and\ \citenamefont
  {Roos}(2012)}]{Blatt2012b}%
  \BibitemOpen
  \bibfield  {author} {\bibinfo {author} {\bibfnamefont {R.}~\bibnamefont
  {Blatt}}\ and\ \bibinfo {author} {\bibfnamefont {C.~F.}\ \bibnamefont
  {Roos}},\ }\href {\doibase 10.1038/nphys2252} {\bibfield  {journal} {\bibinfo
   {journal} {Nat. Phys.}\ }\textbf {\bibinfo {volume} {8}},\ \bibinfo {pages}
  {277} (\bibinfo {year} {2012})}\BibitemShut {NoStop}%
\bibitem [{\citenamefont {{Jurcevic}}\ \emph {et~al.}(2014)\citenamefont
  {{Jurcevic}}, \citenamefont {{Lanyon}}, \citenamefont {{Hauke}},
  \citenamefont {{Hempel}}, \citenamefont {{Zoller}}, \citenamefont {{Blatt}},\
  and\ \citenamefont {{Roos}}}]{Jurcevic2014}%
  \BibitemOpen
  \bibfield  {author} {\bibinfo {author} {\bibfnamefont {P.}~\bibnamefont
  {{Jurcevic}}}, \bibinfo {author} {\bibfnamefont {B.~P.}\ \bibnamefont
  {{Lanyon}}}, \bibinfo {author} {\bibfnamefont {P.}~\bibnamefont {{Hauke}}},
  \bibinfo {author} {\bibfnamefont {C.}~\bibnamefont {{Hempel}}}, \bibinfo
  {author} {\bibfnamefont {P.}~\bibnamefont {{Zoller}}}, \bibinfo {author}
  {\bibfnamefont {R.}~\bibnamefont {{Blatt}}}, \ and\ \bibinfo {author}
  {\bibfnamefont {C.~F.}\ \bibnamefont {{Roos}}},\ }\href {\doibase
  10.1038/nature13461} {\bibfield  {journal} {\bibinfo  {journal} {Nature}\
  }\textbf {\bibinfo {volume} {511}},\ \bibinfo {pages} {202} (\bibinfo {year}
  {2014})}\BibitemShut {NoStop}%
\bibitem [{\citenamefont {{Richerme}}\ \emph {et~al.}(2014)\citenamefont
  {{Richerme}}, \citenamefont {{Gong}}, \citenamefont {{Lee}}, \citenamefont
  {{Senko}}, \citenamefont {{Smith}}, \citenamefont {{Foss-Feig}},
  \citenamefont {{Michalakis}}, \citenamefont {{Gorshkov}},\ and\ \citenamefont
  {{Monroe}}}]{Richerme2014}%
  \BibitemOpen
  \bibfield  {author} {\bibinfo {author} {\bibfnamefont {P.}~\bibnamefont
  {{Richerme}}}, \bibinfo {author} {\bibfnamefont {Z.-X.}\ \bibnamefont
  {{Gong}}}, \bibinfo {author} {\bibfnamefont {A.}~\bibnamefont {{Lee}}},
  \bibinfo {author} {\bibfnamefont {C.}~\bibnamefont {{Senko}}}, \bibinfo
  {author} {\bibfnamefont {J.}~\bibnamefont {{Smith}}}, \bibinfo {author}
  {\bibfnamefont {M.}~\bibnamefont {{Foss-Feig}}}, \bibinfo {author}
  {\bibfnamefont {S.}~\bibnamefont {{Michalakis}}}, \bibinfo {author}
  {\bibfnamefont {A.~V.}\ \bibnamefont {{Gorshkov}}}, \ and\ \bibinfo {author}
  {\bibfnamefont {C.}~\bibnamefont {{Monroe}}},\ }\href {\doibase
  10.1038/nature13450} {\bibfield  {journal} {\bibinfo  {journal} {Nature}\
  }\textbf {\bibinfo {volume} {511}},\ \bibinfo {pages} {198} (\bibinfo {year}
  {2014})}\BibitemShut {NoStop}%
\bibitem [{\citenamefont {Bohnet}\ \emph {et~al.}(2016)\citenamefont {Bohnet},
  \citenamefont {Sawyer}, \citenamefont {Britton}, \citenamefont {Wall},
  \citenamefont {Rey}, \citenamefont {Foss-Feig},\ and\ \citenamefont
  {Bollinger}}]{Bohnet2016a}%
  \BibitemOpen
  \bibfield  {author} {\bibinfo {author} {\bibfnamefont {J.~G.}\ \bibnamefont
  {Bohnet}}, \bibinfo {author} {\bibfnamefont {B.~C.}\ \bibnamefont {Sawyer}},
  \bibinfo {author} {\bibfnamefont {J.~W.}\ \bibnamefont {Britton}}, \bibinfo
  {author} {\bibfnamefont {M.~L.}\ \bibnamefont {Wall}}, \bibinfo {author}
  {\bibfnamefont {A.~M.}\ \bibnamefont {Rey}}, \bibinfo {author} {\bibfnamefont
  {M.}~\bibnamefont {Foss-Feig}}, \ and\ \bibinfo {author} {\bibfnamefont
  {J.~J.}\ \bibnamefont {Bollinger}},\ }\href {\doibase
  10.1126/science.aad9958} {\bibfield  {journal} {\bibinfo  {journal}
  {Science}\ }\textbf {\bibinfo {volume} {352}},\ \bibinfo {pages} {1297}
  (\bibinfo {year} {2016})}\BibitemShut {NoStop}%
\bibitem [{\citenamefont {Glaetzle}\ \emph {et~al.}(2016)\citenamefont
  {Glaetzle}, \citenamefont {van Bijnen}, \citenamefont {Zoller},\ and\
  \citenamefont {Lechner}}]{Glaetzle2016}%
  \BibitemOpen
  \bibfield  {author} {\bibinfo {author} {\bibfnamefont {A.~W.}\ \bibnamefont
  {Glaetzle}}, \bibinfo {author} {\bibfnamefont {R.~M.~W.}\ \bibnamefont {van
  Bijnen}}, \bibinfo {author} {\bibfnamefont {P.}~\bibnamefont {Zoller}}, \
  and\ \bibinfo {author} {\bibfnamefont {W.}~\bibnamefont {Lechner}},\
  }\href@noop {} {\bibfield  {journal} {\bibinfo  {journal} {arXiv:1611.02594}\
  } (\bibinfo {year} {2016})}\BibitemShut {NoStop}%
\bibitem [{\citenamefont {{Glaetzle}}\ \emph {et~al.}(2015)\citenamefont
  {{Glaetzle}}, \citenamefont {{Dalmonte}}, \citenamefont {{Nath}},
  \citenamefont {{Gross}}, \citenamefont {{Bloch}},\ and\ \citenamefont
  {{Zoller}}}]{Glaetzle2015}%
  \BibitemOpen
  \bibfield  {author} {\bibinfo {author} {\bibfnamefont {A.~W.}\ \bibnamefont
  {{Glaetzle}}}, \bibinfo {author} {\bibfnamefont {M.}~\bibnamefont
  {{Dalmonte}}}, \bibinfo {author} {\bibfnamefont {R.}~\bibnamefont {{Nath}}},
  \bibinfo {author} {\bibfnamefont {C.}~\bibnamefont {{Gross}}}, \bibinfo
  {author} {\bibfnamefont {I.}~\bibnamefont {{Bloch}}}, \ and\ \bibinfo
  {author} {\bibfnamefont {P.}~\bibnamefont {{Zoller}}},\ }\href {\doibase
  10.1103/PhysRevLett.114.173002} {\bibfield  {journal} {\bibinfo  {journal}
  {Phys. Rev. Lett.}\ }\textbf {\bibinfo {volume} {114}},\ \bibinfo {pages}
  {173002} (\bibinfo {year} {2015})}\BibitemShut {NoStop}%
\bibitem [{\citenamefont {{van Bijnen}}\ and\ \citenamefont
  {{Pohl}}(2015)}]{vanBijnen2015}%
  \BibitemOpen
  \bibfield  {author} {\bibinfo {author} {\bibfnamefont {R.}~\bibnamefont {{van
  Bijnen}}}\ and\ \bibinfo {author} {\bibfnamefont {T.}~\bibnamefont
  {{Pohl}}},\ }\href {\doibase 10.1103/PhysRevLett.114.243002} {\bibfield
  {journal} {\bibinfo  {journal} {Phys. Rev. Lett.}\ }\textbf {\bibinfo
  {volume} {114}},\ \bibinfo {pages} {243002} (\bibinfo {year}
  {2015})}\BibitemShut {NoStop}%
\bibitem [{\citenamefont {{Henkel}}\ \emph {et~al.}(2010)\citenamefont
  {{Henkel}}, \citenamefont {{Nath}},\ and\ \citenamefont
  {{Pohl}}}]{Henkel2010}%
  \BibitemOpen
  \bibfield  {author} {\bibinfo {author} {\bibfnamefont {N.}~\bibnamefont
  {{Henkel}}}, \bibinfo {author} {\bibfnamefont {R.}~\bibnamefont {{Nath}}}, \
  and\ \bibinfo {author} {\bibfnamefont {T.}~\bibnamefont {{Pohl}}},\ }\href
  {\doibase 10.1103/PhysRevLett.104.195302} {\bibfield  {journal} {\bibinfo
  {journal} {Phys. Rev. Lett.}\ }\textbf {\bibinfo {volume} {104}},\ \bibinfo
  {pages} {195302} (\bibinfo {year} {2010})}\BibitemShut {NoStop}%
\bibitem [{\citenamefont {{Mattioli}}\ \emph {et~al.}(2013)\citenamefont
  {{Mattioli}}, \citenamefont {{Dalmonte}}, \citenamefont {{Lechner}},\ and\
  \citenamefont {{Pupillo}}}]{Mattioli2013}%
  \BibitemOpen
  \bibfield  {author} {\bibinfo {author} {\bibfnamefont {M.}~\bibnamefont
  {{Mattioli}}}, \bibinfo {author} {\bibfnamefont {M.}~\bibnamefont
  {{Dalmonte}}}, \bibinfo {author} {\bibfnamefont {W.}~\bibnamefont
  {{Lechner}}}, \ and\ \bibinfo {author} {\bibfnamefont {G.}~\bibnamefont
  {{Pupillo}}},\ }\href {\doibase 10.1103/PhysRevLett.111.165302} {\bibfield
  {journal} {\bibinfo  {journal} {Phys. Rev. Lett.}\ }\textbf {\bibinfo
  {volume} {111}},\ \bibinfo {pages} {165302} (\bibinfo {year}
  {2013})}\BibitemShut {NoStop}%
\bibitem [{\citenamefont {Gei{\ss}ler}\ \emph {et~al.}(2015)\citenamefont
  {Gei{\ss}ler}, \citenamefont {Vasi{\'c}},\ and\ \citenamefont
  {Hofstetter}}]{Geissler2015}%
  \BibitemOpen
  \bibfield  {author} {\bibinfo {author} {\bibfnamefont {A.}~\bibnamefont
  {Gei{\ss}ler}}, \bibinfo {author} {\bibfnamefont {I.}~\bibnamefont
  {Vasi{\'c}}}, \ and\ \bibinfo {author} {\bibfnamefont {W.}~\bibnamefont
  {Hofstetter}},\ }\href@noop {} {\bibfield  {journal} {\bibinfo  {journal}
  {arXiv:1509.06292}\ } (\bibinfo {year} {2015})}\BibitemShut {NoStop}%
\bibitem [{\citenamefont {Rempe}\ \emph {et~al.}(1987)\citenamefont {Rempe},
  \citenamefont {Walther},\ and\ \citenamefont {Klein}}]{Rempe1987}%
  \BibitemOpen
  \bibfield  {author} {\bibinfo {author} {\bibfnamefont {G.}~\bibnamefont
  {Rempe}}, \bibinfo {author} {\bibfnamefont {H.}~\bibnamefont {Walther}}, \
  and\ \bibinfo {author} {\bibfnamefont {N.}~\bibnamefont {Klein}},\ }\href
  {\doibase 10.1103/PhysRevLett.58.353} {\bibfield  {journal} {\bibinfo
  {journal} {Phys. Rev. Lett.}\ }\textbf {\bibinfo {volume} {58}},\ \bibinfo
  {pages} {353} (\bibinfo {year} {1987})}\BibitemShut {NoStop}%
\bibitem [{\citenamefont {Brune}\ \emph {et~al.}(1996)\citenamefont {Brune},
  \citenamefont {Schmidt-Kaler}, \citenamefont {Maali}, \citenamefont {Dreyer},
  \citenamefont {Hagley}, \citenamefont {Raimond},\ and\ \citenamefont
  {Haroche}}]{Brune1996}%
  \BibitemOpen
  \bibfield  {author} {\bibinfo {author} {\bibfnamefont {M.}~\bibnamefont
  {Brune}}, \bibinfo {author} {\bibfnamefont {F.}~\bibnamefont
  {Schmidt-Kaler}}, \bibinfo {author} {\bibfnamefont {A.}~\bibnamefont
  {Maali}}, \bibinfo {author} {\bibfnamefont {J.}~\bibnamefont {Dreyer}},
  \bibinfo {author} {\bibfnamefont {E.}~\bibnamefont {Hagley}}, \bibinfo
  {author} {\bibfnamefont {J.~M.}\ \bibnamefont {Raimond}}, \ and\ \bibinfo
  {author} {\bibfnamefont {S.}~\bibnamefont {Haroche}},\ }\href {\doibase
  10.1103/PhysRevLett.76.1800} {\bibfield  {journal} {\bibinfo  {journal}
  {Phys. Rev. Lett.}\ }\textbf {\bibinfo {volume} {76}},\ \bibinfo {pages}
  {1800} (\bibinfo {year} {1996})}\BibitemShut {NoStop}%
\bibitem [{\citenamefont {Meekhof}\ \emph {et~al.}(1996)\citenamefont
  {Meekhof}, \citenamefont {Monroe}, \citenamefont {King}, \citenamefont
  {Itano},\ and\ \citenamefont {Wineland}}]{Meekhof1996}%
  \BibitemOpen
  \bibfield  {author} {\bibinfo {author} {\bibfnamefont {D.~M.}\ \bibnamefont
  {Meekhof}}, \bibinfo {author} {\bibfnamefont {C.}~\bibnamefont {Monroe}},
  \bibinfo {author} {\bibfnamefont {B.~E.}\ \bibnamefont {King}}, \bibinfo
  {author} {\bibfnamefont {W.~M.}\ \bibnamefont {Itano}}, \ and\ \bibinfo
  {author} {\bibfnamefont {D.~J.}\ \bibnamefont {Wineland}},\ }\href {\doibase
  10.1103/PhysRevLett.76.1796} {\bibfield  {journal} {\bibinfo  {journal}
  {Phys. Rev. Lett.}\ }\textbf {\bibinfo {volume} {76}},\ \bibinfo {pages}
  {1796} (\bibinfo {year} {1996})}\BibitemShut {NoStop}%
\bibitem [{\citenamefont {Kirchmair}\ \emph {et~al.}(2013)\citenamefont
  {Kirchmair}, \citenamefont {Vlastakis}, \citenamefont {Leghtas},
  \citenamefont {Nigg}, \citenamefont {Paik}, \citenamefont {Ginossar},
  \citenamefont {Mirrahimi}, \citenamefont {Frunzio}, \citenamefont {Girvin},\
  and\ \citenamefont {Schoelkopf}}]{Kirchmair2013}%
  \BibitemOpen
  \bibfield  {author} {\bibinfo {author} {\bibfnamefont {G.}~\bibnamefont
  {Kirchmair}}, \bibinfo {author} {\bibfnamefont {B.}~\bibnamefont
  {Vlastakis}}, \bibinfo {author} {\bibfnamefont {Z.}~\bibnamefont {Leghtas}},
  \bibinfo {author} {\bibfnamefont {S.~E.}\ \bibnamefont {Nigg}}, \bibinfo
  {author} {\bibfnamefont {H.}~\bibnamefont {Paik}}, \bibinfo {author}
  {\bibfnamefont {E.}~\bibnamefont {Ginossar}}, \bibinfo {author}
  {\bibfnamefont {M.}~\bibnamefont {Mirrahimi}}, \bibinfo {author}
  {\bibfnamefont {L.}~\bibnamefont {Frunzio}}, \bibinfo {author} {\bibfnamefont
  {S.~M.}\ \bibnamefont {Girvin}}, \ and\ \bibinfo {author} {\bibfnamefont
  {R.~J.}\ \bibnamefont {Schoelkopf}},\ }\href {\doibase 10.1038/nature11902}
  {\bibfield  {journal} {\bibinfo  {journal} {Nature}\ }\textbf {\bibinfo
  {volume} {495}},\ \bibinfo {pages} {205} (\bibinfo {year}
  {2013})}\BibitemShut {NoStop}%
\bibitem [{\citenamefont {Greiner}\ \emph {et~al.}(2002)\citenamefont
  {Greiner}, \citenamefont {Mandel}, \citenamefont {H{\"a}nsch},\ and\
  \citenamefont {Bloch}}]{Greiner2002}%
  \BibitemOpen
  \bibfield  {author} {\bibinfo {author} {\bibfnamefont {M.}~\bibnamefont
  {Greiner}}, \bibinfo {author} {\bibfnamefont {O.}~\bibnamefont {Mandel}},
  \bibinfo {author} {\bibfnamefont {T.~W.}\ \bibnamefont {H{\"a}nsch}}, \ and\
  \bibinfo {author} {\bibfnamefont {I.}~\bibnamefont {Bloch}},\ }\href
  {\doibase 10.1038/nature00968} {\bibfield  {journal} {\bibinfo  {journal}
  {Nature}\ }\textbf {\bibinfo {volume} {419}},\ \bibinfo {pages} {51}
  (\bibinfo {year} {2002})}\BibitemShut {NoStop}%
\bibitem [{\citenamefont {Will}\ \emph {et~al.}(2010)\citenamefont {Will},
  \citenamefont {Best}, \citenamefont {Schneider}, \citenamefont
  {Hackerm{\"u}ller}, \citenamefont {L{\"u}hmann},\ and\ \citenamefont
  {Bloch}}]{Will2010}%
  \BibitemOpen
  \bibfield  {author} {\bibinfo {author} {\bibfnamefont {S.}~\bibnamefont
  {Will}}, \bibinfo {author} {\bibfnamefont {T.}~\bibnamefont {Best}}, \bibinfo
  {author} {\bibfnamefont {U.}~\bibnamefont {Schneider}}, \bibinfo {author}
  {\bibfnamefont {L.}~\bibnamefont {Hackerm{\"u}ller}}, \bibinfo {author}
  {\bibfnamefont {D.-S.}\ \bibnamefont {L{\"u}hmann}}, \ and\ \bibinfo {author}
  {\bibfnamefont {I.}~\bibnamefont {Bloch}},\ }\href {\doibase
  10.1038/nature09036} {\bibfield  {journal} {\bibinfo  {journal} {Nature}\
  }\textbf {\bibinfo {volume} {465}},\ \bibinfo {pages} {197} (\bibinfo {year}
  {2010})}\BibitemShut {NoStop}%
\bibitem [{\citenamefont {{Yan}}\ \emph {et~al.}(2013)\citenamefont {{Yan}},
  \citenamefont {{Moses}}, \citenamefont {{Gadway}}, \citenamefont {{Covey}},
  \citenamefont {{Hazzard}}, \citenamefont {{Rey}}, \citenamefont {{Jin}},\
  and\ \citenamefont {{Ye}}}]{Yan2013}%
  \BibitemOpen
  \bibfield  {author} {\bibinfo {author} {\bibfnamefont {B.}~\bibnamefont
  {{Yan}}}, \bibinfo {author} {\bibfnamefont {S.~A.}\ \bibnamefont {{Moses}}},
  \bibinfo {author} {\bibfnamefont {B.}~\bibnamefont {{Gadway}}}, \bibinfo
  {author} {\bibfnamefont {J.~P.}\ \bibnamefont {{Covey}}}, \bibinfo {author}
  {\bibfnamefont {K.~R.~A.}\ \bibnamefont {{Hazzard}}}, \bibinfo {author}
  {\bibfnamefont {A.~M.}\ \bibnamefont {{Rey}}}, \bibinfo {author}
  {\bibfnamefont {D.~S.}\ \bibnamefont {{Jin}}}, \ and\ \bibinfo {author}
  {\bibfnamefont {J.}~\bibnamefont {{Ye}}},\ }\href {\doibase
  10.1038/nature12483} {\bibfield  {journal} {\bibinfo  {journal} {Nature}\
  }\textbf {\bibinfo {volume} {501}},\ \bibinfo {pages} {521} (\bibinfo {year}
  {2013})}\BibitemShut {NoStop}%
\bibitem [{\citenamefont {{de Paz}}\ \emph {et~al.}(2013)\citenamefont {{de
  Paz}}, \citenamefont {Sharma}, \citenamefont {Chotia}, \citenamefont
  {Mar{\'e}chal}, \citenamefont {Huckans}, \citenamefont {Pedri}, \citenamefont
  {Santos}, \citenamefont {Gorceix}, \citenamefont {Vernac},\ and\
  \citenamefont {Laburthe-Tolra}}]{dePaz2013a}%
  \BibitemOpen
  \bibfield  {author} {\bibinfo {author} {\bibfnamefont {A.}~\bibnamefont {{de
  Paz}}}, \bibinfo {author} {\bibfnamefont {A.}~\bibnamefont {Sharma}},
  \bibinfo {author} {\bibfnamefont {A.}~\bibnamefont {Chotia}}, \bibinfo
  {author} {\bibfnamefont {E.}~\bibnamefont {Mar{\'e}chal}}, \bibinfo {author}
  {\bibfnamefont {J.~H.}\ \bibnamefont {Huckans}}, \bibinfo {author}
  {\bibfnamefont {P.}~\bibnamefont {Pedri}}, \bibinfo {author} {\bibfnamefont
  {L.}~\bibnamefont {Santos}}, \bibinfo {author} {\bibfnamefont
  {O.}~\bibnamefont {Gorceix}}, \bibinfo {author} {\bibfnamefont
  {L.}~\bibnamefont {Vernac}}, \ and\ \bibinfo {author} {\bibfnamefont
  {B.}~\bibnamefont {Laburthe-Tolra}},\ }\href {\doibase
  10.1103/PhysRevLett.111.185305} {\bibfield  {journal} {\bibinfo  {journal}
  {Phys. Rev. Lett.}\ }\textbf {\bibinfo {volume} {111}},\ \bibinfo {pages}
  {185305} (\bibinfo {year} {2013})}\BibitemShut {NoStop}%
\bibitem [{\citenamefont {{Schau{\ss}}}\ \emph {et~al.}(2015)\citenamefont
  {{Schau{\ss}}}, \citenamefont {{Zeiher}}, \citenamefont {{Fukuhara}},
  \citenamefont {{Hild}}, \citenamefont {{Cheneau}}, \citenamefont
  {{Macr{\`\i}}}, \citenamefont {{Pohl}}, \citenamefont {{Bloch}},\ and\
  \citenamefont {{Gross}}}]{Schauss2015}%
  \BibitemOpen
  \bibfield  {author} {\bibinfo {author} {\bibfnamefont {P.}~\bibnamefont
  {{Schau{\ss}}}}, \bibinfo {author} {\bibfnamefont {J.}~\bibnamefont
  {{Zeiher}}}, \bibinfo {author} {\bibfnamefont {T.}~\bibnamefont
  {{Fukuhara}}}, \bibinfo {author} {\bibfnamefont {S.}~\bibnamefont {{Hild}}},
  \bibinfo {author} {\bibfnamefont {M.}~\bibnamefont {{Cheneau}}}, \bibinfo
  {author} {\bibfnamefont {T.}~\bibnamefont {{Macr{\`\i}}}}, \bibinfo {author}
  {\bibfnamefont {T.}~\bibnamefont {{Pohl}}}, \bibinfo {author} {\bibfnamefont
  {I.}~\bibnamefont {{Bloch}}}, \ and\ \bibinfo {author} {\bibfnamefont
  {C.}~\bibnamefont {{Gross}}},\ }\href {\doibase 10.1126/science.1258351}
  {\bibfield  {journal} {\bibinfo  {journal} {Science}\ }\textbf {\bibinfo
  {volume} {347}},\ \bibinfo {pages} {1455} (\bibinfo {year}
  {2015})}\BibitemShut {NoStop}%
\bibitem [{\citenamefont {Labuhn}\ \emph {et~al.}(2016)\citenamefont {Labuhn},
  \citenamefont {Barredo}, \citenamefont {Ravets}, \citenamefont {{de
  L{\'e}s{\'e}leuc}}, \citenamefont {Macr{\`\i}}, \citenamefont {Lahaye},\ and\
  \citenamefont {Browaeys}}]{Labuhn2016a}%
  \BibitemOpen
  \bibfield  {author} {\bibinfo {author} {\bibfnamefont {H.}~\bibnamefont
  {Labuhn}}, \bibinfo {author} {\bibfnamefont {D.}~\bibnamefont {Barredo}},
  \bibinfo {author} {\bibfnamefont {S.}~\bibnamefont {Ravets}}, \bibinfo
  {author} {\bibfnamefont {S.}~\bibnamefont {{de L{\'e}s{\'e}leuc}}}, \bibinfo
  {author} {\bibfnamefont {T.}~\bibnamefont {Macr{\`\i}}}, \bibinfo {author}
  {\bibfnamefont {T.}~\bibnamefont {Lahaye}}, \ and\ \bibinfo {author}
  {\bibfnamefont {A.}~\bibnamefont {Browaeys}},\ }\href {\doibase
  10.1038/nature18274} {\bibfield  {journal} {\bibinfo  {journal} {Nature}\
  }\textbf {\bibinfo {volume} {534}},\ \bibinfo {pages} {667} (\bibinfo {year}
  {2016})}\BibitemShut {NoStop}%
\bibitem [{\citenamefont {Jau}\ \emph {et~al.}(2016)\citenamefont {Jau},
  \citenamefont {Hankin}, \citenamefont {Keating}, \citenamefont {Deutsch},\
  and\ \citenamefont {Biedermann}}]{Jau2016}%
  \BibitemOpen
  \bibfield  {author} {\bibinfo {author} {\bibfnamefont {Y.-Y.}\ \bibnamefont
  {Jau}}, \bibinfo {author} {\bibfnamefont {A.~M.}\ \bibnamefont {Hankin}},
  \bibinfo {author} {\bibfnamefont {T.}~\bibnamefont {Keating}}, \bibinfo
  {author} {\bibfnamefont {I.~H.}\ \bibnamefont {Deutsch}}, \ and\ \bibinfo
  {author} {\bibfnamefont {G.~W.}\ \bibnamefont {Biedermann}},\ }\href
  {\doibase 10.1038/nphys3487} {\bibfield  {journal} {\bibinfo  {journal} {Nat.
  Phys.}\ }\textbf {\bibinfo {volume} {12}},\ \bibinfo {pages} {71} (\bibinfo
  {year} {2016})}\BibitemShut {NoStop}%
\bibitem [{\citenamefont {Zeiher}\ \emph {et~al.}(2016)\citenamefont {Zeiher},
  \citenamefont {{van Bijnen}}, \citenamefont {Schau\ss{}}, \citenamefont
  {Hild}, \citenamefont {Choi}, \citenamefont {Pohl}, \citenamefont {Bloch},\
  and\ \citenamefont {Gross}}]{Zeiher2016a}%
  \BibitemOpen
  \bibfield  {author} {\bibinfo {author} {\bibfnamefont {J.}~\bibnamefont
  {Zeiher}}, \bibinfo {author} {\bibfnamefont {R.}~\bibnamefont {{van
  Bijnen}}}, \bibinfo {author} {\bibfnamefont {P.}~\bibnamefont {Schau\ss{}}},
  \bibinfo {author} {\bibfnamefont {S.}~\bibnamefont {Hild}}, \bibinfo {author}
  {\bibfnamefont {J.-y.}\ \bibnamefont {Choi}}, \bibinfo {author}
  {\bibfnamefont {T.}~\bibnamefont {Pohl}}, \bibinfo {author} {\bibfnamefont
  {I.}~\bibnamefont {Bloch}}, \ and\ \bibinfo {author} {\bibfnamefont
  {C.}~\bibnamefont {Gross}},\ }\href {\doibase 10.1038/nphys3835} {\bibfield
  {journal} {\bibinfo  {journal} {Nat. Phys.}\ }\textbf {\bibinfo {volume}
  {12}},\ \bibinfo {pages} {1095} (\bibinfo {year} {2016})}\BibitemShut
  {NoStop}%
\bibitem [{\citenamefont {{Barredo}}\ \emph {et~al.}(2015)\citenamefont
  {{Barredo}}, \citenamefont {{Labuhn}}, \citenamefont {{Ravets}},
  \citenamefont {{Lahaye}}, \citenamefont {{Browaeys}},\ and\ \citenamefont
  {{Adams}}}]{Barredo2015a}%
  \BibitemOpen
  \bibfield  {author} {\bibinfo {author} {\bibfnamefont {D.}~\bibnamefont
  {{Barredo}}}, \bibinfo {author} {\bibfnamefont {H.}~\bibnamefont {{Labuhn}}},
  \bibinfo {author} {\bibfnamefont {S.}~\bibnamefont {{Ravets}}}, \bibinfo
  {author} {\bibfnamefont {T.}~\bibnamefont {{Lahaye}}}, \bibinfo {author}
  {\bibfnamefont {A.}~\bibnamefont {{Browaeys}}}, \ and\ \bibinfo {author}
  {\bibfnamefont {C.~S.}\ \bibnamefont {{Adams}}},\ }\href {\doibase
  10.1103/PhysRevLett.114.113002} {\bibfield  {journal} {\bibinfo  {journal}
  {Phys. Rev. Lett.}\ }\textbf {\bibinfo {volume} {114}},\ \bibinfo {pages}
  {113002} (\bibinfo {year} {2015})}\BibitemShut {NoStop}%
\bibitem [{\citenamefont {{Santos}}\ \emph {et~al.}(2000)\citenamefont
  {{Santos}}, \citenamefont {{Shlyapnikov}}, \citenamefont {{Zoller}},\ and\
  \citenamefont {{Lewenstein}}}]{Santos2000}%
  \BibitemOpen
  \bibfield  {author} {\bibinfo {author} {\bibfnamefont {L.}~\bibnamefont
  {{Santos}}}, \bibinfo {author} {\bibfnamefont {G.~V.}\ \bibnamefont
  {{Shlyapnikov}}}, \bibinfo {author} {\bibfnamefont {P.}~\bibnamefont
  {{Zoller}}}, \ and\ \bibinfo {author} {\bibfnamefont {M.}~\bibnamefont
  {{Lewenstein}}},\ }\href {\doibase 10.1103/PhysRevLett.85.1791} {\bibfield
  {journal} {\bibinfo  {journal} {Phys. Rev. Lett.}\ }\textbf {\bibinfo
  {volume} {85}},\ \bibinfo {pages} {1791} (\bibinfo {year}
  {2000})}\BibitemShut {NoStop}%
\bibitem [{\citenamefont {{Bouchoule}}\ and\ \citenamefont
  {{M{\o}lmer}}(2002)}]{Bouchoule2002}%
  \BibitemOpen
  \bibfield  {author} {\bibinfo {author} {\bibfnamefont {I.}~\bibnamefont
  {{Bouchoule}}}\ and\ \bibinfo {author} {\bibfnamefont {K.}~\bibnamefont
  {{M{\o}lmer}}},\ }\href {\doibase 10.1103/PhysRevA.65.041803} {\bibfield
  {journal} {\bibinfo  {journal} {Phys. Rev. A}\ }\textbf {\bibinfo {volume}
  {65}},\ \bibinfo {pages} {041803} (\bibinfo {year} {2002})}\BibitemShut
  {NoStop}%
\bibitem [{SI()}]{SI}%
  \BibitemOpen
  \href@noop {} {}\bibinfo {note} {{see Supplementary Information}}\BibitemShut
  {NoStop}%
\bibitem [{\citenamefont {Radcliffe}(1971)}]{Radcliffe1971}%
  \BibitemOpen
  \bibfield  {author} {\bibinfo {author} {\bibfnamefont {J.~M.}\ \bibnamefont
  {Radcliffe}},\ }\href {\doibase 10.1088/0305-4470/4/3/009} {\bibfield
  {journal} {\bibinfo  {journal} {J. Phys. A: Gen. Phys.}\ }\textbf {\bibinfo
  {volume} {4}},\ \bibinfo {pages} {313} (\bibinfo {year} {1971})}\BibitemShut
  {NoStop}%
\bibitem [{\citenamefont {Hazzard}\ \emph {et~al.}(2014)\citenamefont
  {Hazzard}, \citenamefont {{van den Worm}}, \citenamefont {Foss-Feig},
  \citenamefont {Manmana}, \citenamefont {Dalla~Torre}, \citenamefont {Pfau},
  \citenamefont {Kastner},\ and\ \citenamefont {Rey}}]{Hazzard2014}%
  \BibitemOpen
  \bibfield  {author} {\bibinfo {author} {\bibfnamefont {K.~R.~A.}\
  \bibnamefont {Hazzard}}, \bibinfo {author} {\bibfnamefont {M.}~\bibnamefont
  {{van den Worm}}}, \bibinfo {author} {\bibfnamefont {M.}~\bibnamefont
  {Foss-Feig}}, \bibinfo {author} {\bibfnamefont {S.~R.}\ \bibnamefont
  {Manmana}}, \bibinfo {author} {\bibfnamefont {E.~G.}\ \bibnamefont
  {Dalla~Torre}}, \bibinfo {author} {\bibfnamefont {T.}~\bibnamefont {Pfau}},
  \bibinfo {author} {\bibfnamefont {M.}~\bibnamefont {Kastner}}, \ and\
  \bibinfo {author} {\bibfnamefont {A.~M.}\ \bibnamefont {Rey}},\ }\href
  {\doibase 10.1103/PhysRevA.90.063622} {\bibfield  {journal} {\bibinfo
  {journal} {Phys. Rev. A}\ }\textbf {\bibinfo {volume} {90}},\ \bibinfo
  {pages} {063622} (\bibinfo {year} {2014})}\BibitemShut {NoStop}%
\bibitem [{\citenamefont {Macr{\`\i}}\ \emph {et~al.}(2016)\citenamefont
  {Macr{\`\i}}, \citenamefont {Smerzi},\ and\ \citenamefont
  {Pezz{\`e}}}]{Macri2016a}%
  \BibitemOpen
  \bibfield  {author} {\bibinfo {author} {\bibfnamefont {T.}~\bibnamefont
  {Macr{\`\i}}}, \bibinfo {author} {\bibfnamefont {A.}~\bibnamefont {Smerzi}},
  \ and\ \bibinfo {author} {\bibfnamefont {L.}~\bibnamefont {Pezz{\`e}}},\
  }\href {\doibase 10.1103/PhysRevA.94.010102} {\bibfield  {journal} {\bibinfo
  {journal} {Phys. Rev. A}\ }\textbf {\bibinfo {volume} {94}},\ \bibinfo
  {pages} {010102} (\bibinfo {year} {2016})}\BibitemShut {NoStop}%
\bibitem [{\citenamefont {Foss-Feig}\ \emph {et~al.}(2016)\citenamefont
  {Foss-Feig}, \citenamefont {Gong}, \citenamefont {Gorshkov},\ and\
  \citenamefont {Clark}}]{Foss-Feig2016}%
  \BibitemOpen
  \bibfield  {author} {\bibinfo {author} {\bibfnamefont {M.}~\bibnamefont
  {Foss-Feig}}, \bibinfo {author} {\bibfnamefont {Z.-X.}\ \bibnamefont {Gong}},
  \bibinfo {author} {\bibfnamefont {A.~V.}\ \bibnamefont {Gorshkov}}, \ and\
  \bibinfo {author} {\bibfnamefont {C.~W.}\ \bibnamefont {Clark}},\ }\href@noop
  {} {\bibfield  {journal} {\bibinfo  {journal} {arXiv:1612.07805}\ } (\bibinfo
  {year} {2016})}\BibitemShut {NoStop}%
\bibitem [{\citenamefont {Raussendorf}\ and\ \citenamefont
  {Briegel}(2001)}]{Raussendorf2001}%
  \BibitemOpen
  \bibfield  {author} {\bibinfo {author} {\bibfnamefont {R.}~\bibnamefont
  {Raussendorf}}\ and\ \bibinfo {author} {\bibfnamefont {H.~J.}\ \bibnamefont
  {Briegel}},\ }\href {\doibase 10.1103/PhysRevLett.86.5188} {\bibfield
  {journal} {\bibinfo  {journal} {Phys. Rev. Lett.}\ }\textbf {\bibinfo
  {volume} {86}},\ \bibinfo {pages} {5188} (\bibinfo {year}
  {2001})}\BibitemShut {NoStop}%
\bibitem [{\citenamefont {Foss-Feig}\ \emph {et~al.}(2013)\citenamefont
  {Foss-Feig}, \citenamefont {Hazzard}, \citenamefont {Bollinger},\ and\
  \citenamefont {Rey}}]{Foss-Feig2013}%
  \BibitemOpen
  \bibfield  {author} {\bibinfo {author} {\bibfnamefont {M.}~\bibnamefont
  {Foss-Feig}}, \bibinfo {author} {\bibfnamefont {K.~R.~A.}\ \bibnamefont
  {Hazzard}}, \bibinfo {author} {\bibfnamefont {J.~J.}\ \bibnamefont
  {Bollinger}}, \ and\ \bibinfo {author} {\bibfnamefont {A.~M.}\ \bibnamefont
  {Rey}},\ }\href {\doibase 10.1103/PhysRevA.87.042101} {\bibfield  {journal}
  {\bibinfo  {journal} {Phys. Rev. A}\ }\textbf {\bibinfo {volume} {87}},\
  \bibinfo {pages} {042101} (\bibinfo {year} {2013})}\BibitemShut {NoStop}%
\bibitem [{\citenamefont {Schachenmayer}\ \emph {et~al.}(2013)\citenamefont
  {Schachenmayer}, \citenamefont {Lanyon}, \citenamefont {Roos},\ and\
  \citenamefont {Daley}}]{Schachenmayer2013}%
  \BibitemOpen
  \bibfield  {author} {\bibinfo {author} {\bibfnamefont {J.}~\bibnamefont
  {Schachenmayer}}, \bibinfo {author} {\bibfnamefont {B.~P.}\ \bibnamefont
  {Lanyon}}, \bibinfo {author} {\bibfnamefont {C.~F.}\ \bibnamefont {Roos}}, \
  and\ \bibinfo {author} {\bibfnamefont {A.~J.}\ \bibnamefont {Daley}},\ }\href
  {\doibase 10.1103/PhysRevX.3.031015} {\bibfield  {journal} {\bibinfo
  {journal} {Phys. Rev. X}\ }\textbf {\bibinfo {volume} {3}},\ \bibinfo {pages}
  {031015} (\bibinfo {year} {2013})}\BibitemShut {NoStop}%
\bibitem [{\citenamefont {Weitenberg}\ \emph {et~al.}(2011)\citenamefont
  {Weitenberg}, \citenamefont {Endres}, \citenamefont {Sherson}, \citenamefont
  {Cheneau}, \citenamefont {Schau\ss{}}, \citenamefont {Fukuhara},
  \citenamefont {Bloch},\ and\ \citenamefont {Kuhr}}]{Weitenberg2011a}%
  \BibitemOpen
  \bibfield  {author} {\bibinfo {author} {\bibfnamefont {C.}~\bibnamefont
  {Weitenberg}}, \bibinfo {author} {\bibfnamefont {M.}~\bibnamefont {Endres}},
  \bibinfo {author} {\bibfnamefont {J.~F.}\ \bibnamefont {Sherson}}, \bibinfo
  {author} {\bibfnamefont {M.}~\bibnamefont {Cheneau}}, \bibinfo {author}
  {\bibfnamefont {P.}~\bibnamefont {Schau\ss{}}}, \bibinfo {author}
  {\bibfnamefont {T.}~\bibnamefont {Fukuhara}}, \bibinfo {author}
  {\bibfnamefont {I.}~\bibnamefont {Bloch}}, \ and\ \bibinfo {author}
  {\bibfnamefont {S.}~\bibnamefont {Kuhr}},\ }\href {\doibase
  10.1038/nature09827} {\bibfield  {journal} {\bibinfo  {journal} {Nature}\
  }\textbf {\bibinfo {volume} {471}},\ \bibinfo {pages} {319} (\bibinfo {year}
  {2011})}\BibitemShut {NoStop}%
\bibitem [{\citenamefont {{Fukuhara}}\ \emph {et~al.}(2013)\citenamefont
  {{Fukuhara}}, \citenamefont {{Kantian}}, \citenamefont {{Endres}},
  \citenamefont {{Cheneau}}, \citenamefont {{Schau{\ss}}}, \citenamefont
  {{Hild}}, \citenamefont {{Bellem}}, \citenamefont {{Schollw{\"o}ck}},
  \citenamefont {{Giamarchi}}, \citenamefont {{Gross}}, \citenamefont
  {{Bloch}},\ and\ \citenamefont {{Kuhr}}}]{Fukuhara2013}%
  \BibitemOpen
  \bibfield  {author} {\bibinfo {author} {\bibfnamefont {T.}~\bibnamefont
  {{Fukuhara}}}, \bibinfo {author} {\bibfnamefont {A.}~\bibnamefont
  {{Kantian}}}, \bibinfo {author} {\bibfnamefont {M.}~\bibnamefont {{Endres}}},
  \bibinfo {author} {\bibfnamefont {M.}~\bibnamefont {{Cheneau}}}, \bibinfo
  {author} {\bibfnamefont {P.}~\bibnamefont {{Schau{\ss}}}}, \bibinfo {author}
  {\bibfnamefont {S.}~\bibnamefont {{Hild}}}, \bibinfo {author} {\bibfnamefont
  {D.}~\bibnamefont {{Bellem}}}, \bibinfo {author} {\bibfnamefont
  {U.}~\bibnamefont {{Schollw{\"o}ck}}}, \bibinfo {author} {\bibfnamefont
  {T.}~\bibnamefont {{Giamarchi}}}, \bibinfo {author} {\bibfnamefont
  {C.}~\bibnamefont {{Gross}}}, \bibinfo {author} {\bibfnamefont
  {I.}~\bibnamefont {{Bloch}}}, \ and\ \bibinfo {author} {\bibfnamefont
  {S.}~\bibnamefont {{Kuhr}}},\ }\href {\doibase 10.1038/nphys2561} {\bibfield
  {journal} {\bibinfo  {journal} {Nat. Phys.}\ }\textbf {\bibinfo {volume}
  {9}},\ \bibinfo {pages} {235} (\bibinfo {year} {2013})}\BibitemShut {NoStop}%
\bibitem [{\citenamefont {Fukuhara}\ \emph {et~al.}(2015)\citenamefont
  {Fukuhara}, \citenamefont {Hild}, \citenamefont {Zeiher}, \citenamefont
  {Schau\ss{}}, \citenamefont {Bloch}, \citenamefont {Endres},\ and\
  \citenamefont {Gross}}]{Fukuhara2015}%
  \BibitemOpen
  \bibfield  {author} {\bibinfo {author} {\bibfnamefont {T.}~\bibnamefont
  {Fukuhara}}, \bibinfo {author} {\bibfnamefont {S.}~\bibnamefont {Hild}},
  \bibinfo {author} {\bibfnamefont {J.}~\bibnamefont {Zeiher}}, \bibinfo
  {author} {\bibfnamefont {P.}~\bibnamefont {Schau\ss{}}}, \bibinfo {author}
  {\bibfnamefont {I.}~\bibnamefont {Bloch}}, \bibinfo {author} {\bibfnamefont
  {M.}~\bibnamefont {Endres}}, \ and\ \bibinfo {author} {\bibfnamefont
  {C.}~\bibnamefont {Gross}},\ }\href {\doibase 10.1103/PhysRevLett.115.035302}
  {\bibfield  {journal} {\bibinfo  {journal} {Phys. Rev. Lett.}\ }\textbf
  {\bibinfo {volume} {115}},\ \bibinfo {pages} {035302} (\bibinfo {year}
  {2015})}\BibitemShut {NoStop}%
\bibitem [{\citenamefont {Cheneau}\ \emph {et~al.}(2012)\citenamefont
  {Cheneau}, \citenamefont {Barmettler}, \citenamefont {Poletti}, \citenamefont
  {Endres}, \citenamefont {Schau{\ss}}, \citenamefont {Fukuhara}, \citenamefont
  {Gross}, \citenamefont {Bloch}, \citenamefont {Kollath},\ and\ \citenamefont
  {Kuhr}}]{Cheneau2012}%
  \BibitemOpen
  \bibfield  {author} {\bibinfo {author} {\bibfnamefont {M.}~\bibnamefont
  {Cheneau}}, \bibinfo {author} {\bibfnamefont {P.}~\bibnamefont {Barmettler}},
  \bibinfo {author} {\bibfnamefont {D.}~\bibnamefont {Poletti}}, \bibinfo
  {author} {\bibfnamefont {M.}~\bibnamefont {Endres}}, \bibinfo {author}
  {\bibfnamefont {P.}~\bibnamefont {Schau{\ss}}}, \bibinfo {author}
  {\bibfnamefont {T.}~\bibnamefont {Fukuhara}}, \bibinfo {author}
  {\bibfnamefont {C.}~\bibnamefont {Gross}}, \bibinfo {author} {\bibfnamefont
  {I.}~\bibnamefont {Bloch}}, \bibinfo {author} {\bibfnamefont
  {C.}~\bibnamefont {Kollath}}, \ and\ \bibinfo {author} {\bibfnamefont
  {S.}~\bibnamefont {Kuhr}},\ }\href {\doibase 10.1038/nature10748} {\bibfield
  {journal} {\bibinfo  {journal} {Nature}\ }\textbf {\bibinfo {volume} {481}},\
  \bibinfo {pages} {484} (\bibinfo {year} {2012})}\BibitemShut {NoStop}%
\bibitem [{\citenamefont {Breitenbach}\ \emph {et~al.}(1997)\citenamefont
  {Breitenbach}, \citenamefont {Schiller},\ and\ \citenamefont
  {Mlynek}}]{Breitenbach1997}%
  \BibitemOpen
  \bibfield  {author} {\bibinfo {author} {\bibfnamefont {G.}~\bibnamefont
  {Breitenbach}}, \bibinfo {author} {\bibfnamefont {S.}~\bibnamefont
  {Schiller}}, \ and\ \bibinfo {author} {\bibfnamefont {J.}~\bibnamefont
  {Mlynek}},\ }\href {\doibase 10.1038/387471a0} {\bibfield  {journal}
  {\bibinfo  {journal} {Nature}\ }\textbf {\bibinfo {volume} {387}},\ \bibinfo
  {pages} {471} (\bibinfo {year} {1997})}\BibitemShut {NoStop}%
\bibitem [{\citenamefont {Schleich}\ and\ \citenamefont
  {Wheeler}(1987)}]{Schleich1987}%
  \BibitemOpen
  \bibfield  {author} {\bibinfo {author} {\bibfnamefont {W.}~\bibnamefont
  {Schleich}}\ and\ \bibinfo {author} {\bibfnamefont {J.~A.}\ \bibnamefont
  {Wheeler}},\ }\href {\doibase 10.1038/326574a0} {\bibfield  {journal}
  {\bibinfo  {journal} {Nature}\ }\textbf {\bibinfo {volume} {326}},\ \bibinfo
  {pages} {574} (\bibinfo {year} {1987})}\BibitemShut {NoStop}%
\bibitem [{\citenamefont {Zhang}\ \emph {et~al.}(2017)\citenamefont {Zhang},
  \citenamefont {Hess}, \citenamefont {Kyprianidis}, \citenamefont {Becker},
  \citenamefont {Lee}, \citenamefont {Smith}, \citenamefont {Pagano},
  \citenamefont {Potirniche}, \citenamefont {Potter}, \citenamefont
  {Vishwanath}, \citenamefont {Yao},\ and\ \citenamefont {Monroe}}]{Zhang2017}%
  \BibitemOpen
  \bibfield  {author} {\bibinfo {author} {\bibfnamefont {J.}~\bibnamefont
  {Zhang}}, \bibinfo {author} {\bibfnamefont {P.~W.}\ \bibnamefont {Hess}},
  \bibinfo {author} {\bibfnamefont {A.}~\bibnamefont {Kyprianidis}}, \bibinfo
  {author} {\bibfnamefont {P.}~\bibnamefont {Becker}}, \bibinfo {author}
  {\bibfnamefont {A.}~\bibnamefont {Lee}}, \bibinfo {author} {\bibfnamefont
  {J.}~\bibnamefont {Smith}}, \bibinfo {author} {\bibfnamefont
  {G.}~\bibnamefont {Pagano}}, \bibinfo {author} {\bibfnamefont {I.-D.}\
  \bibnamefont {Potirniche}}, \bibinfo {author} {\bibfnamefont {A.~C.}\
  \bibnamefont {Potter}}, \bibinfo {author} {\bibfnamefont {A.}~\bibnamefont
  {Vishwanath}}, \bibinfo {author} {\bibfnamefont {N.~Y.}\ \bibnamefont {Yao}},
  \ and\ \bibinfo {author} {\bibfnamefont {C.}~\bibnamefont {Monroe}},\ }\href
  {\doibase 10.1038/nature21413} {\bibfield  {journal} {\bibinfo  {journal}
  {Nature}\ }\textbf {\bibinfo {volume} {543}},\ \bibinfo {pages} {217}
  (\bibinfo {year} {2017})}\BibitemShut {NoStop}%
\bibitem [{\citenamefont {{Khemani}}\ \emph {et~al.}(2016)\citenamefont
  {{Khemani}}, \citenamefont {{Lazarides}}, \citenamefont {{Moessner}},\ and\
  \citenamefont {{Sondhi}}}]{Khemani2016b}%
  \BibitemOpen
  \bibfield  {author} {\bibinfo {author} {\bibfnamefont {V.}~\bibnamefont
  {{Khemani}}}, \bibinfo {author} {\bibfnamefont {A.}~\bibnamefont
  {{Lazarides}}}, \bibinfo {author} {\bibfnamefont {R.}~\bibnamefont
  {{Moessner}}}, \ and\ \bibinfo {author} {\bibfnamefont {S.~L.}\ \bibnamefont
  {{Sondhi}}},\ }\href {\doibase 10.1103/PhysRevLett.116.250401} {\bibfield
  {journal} {\bibinfo  {journal} {Phys. Rev. Lett.}\ }\textbf {\bibinfo
  {volume} {116}},\ \bibinfo {pages} {250401} (\bibinfo {year}
  {2016})}\BibitemShut {NoStop}%
\bibitem [{\citenamefont {Potirniche}\ \emph {et~al.}(2016)\citenamefont
  {Potirniche}, \citenamefont {Potter}, \citenamefont {Schleier-Smith},
  \citenamefont {Vishwanath},\ and\ \citenamefont {Yao}}]{Potirniche2016a}%
  \BibitemOpen
  \bibfield  {author} {\bibinfo {author} {\bibfnamefont {I.-D.}\ \bibnamefont
  {Potirniche}}, \bibinfo {author} {\bibfnamefont {A.~C.}\ \bibnamefont
  {Potter}}, \bibinfo {author} {\bibfnamefont {M.}~\bibnamefont
  {Schleier-Smith}}, \bibinfo {author} {\bibfnamefont {A.}~\bibnamefont
  {Vishwanath}}, \ and\ \bibinfo {author} {\bibfnamefont {N.~Y.}\ \bibnamefont
  {Yao}},\ }\href@noop {} {\bibfield  {journal} {\bibinfo  {journal}
  {arXiv:1610.07611}\ } (\bibinfo {year} {2016})}\BibitemShut {NoStop}%
\bibitem [{\citenamefont {Else}\ \emph {et~al.}(2016)\citenamefont {Else},
  \citenamefont {Bauer},\ and\ \citenamefont {Nayak}}]{Else2016c}%
  \BibitemOpen
  \bibfield  {author} {\bibinfo {author} {\bibfnamefont {D.~V.}\ \bibnamefont
  {Else}}, \bibinfo {author} {\bibfnamefont {B.}~\bibnamefont {Bauer}}, \ and\
  \bibinfo {author} {\bibfnamefont {C.}~\bibnamefont {Nayak}},\ }\href
  {\doibase 10.1103/PhysRevLett.117.090402} {\bibfield  {journal} {\bibinfo
  {journal} {Phys. Rev. Lett.}\ }\textbf {\bibinfo {volume} {117}},\ \bibinfo
  {pages} {090402} (\bibinfo {year} {2016})}\BibitemShut {NoStop}%
\bibitem [{\citenamefont {Choi}\ \emph {et~al.}(2017)\citenamefont {Choi},
  \citenamefont {Choi}, \citenamefont {Landig}, \citenamefont {Kucsko},
  \citenamefont {Zhou}, \citenamefont {Isoya}, \citenamefont {Jelezko},
  \citenamefont {Onoda}, \citenamefont {Sumiya}, \citenamefont {Khemani},
  \citenamefont {{von Keyserlingk}}, \citenamefont {Yao}, \citenamefont
  {Demler},\ and\ \citenamefont {Lukin}}]{Choi2017}%
  \BibitemOpen
  \bibfield  {author} {\bibinfo {author} {\bibfnamefont {S.}~\bibnamefont
  {Choi}}, \bibinfo {author} {\bibfnamefont {J.}~\bibnamefont {Choi}}, \bibinfo
  {author} {\bibfnamefont {R.}~\bibnamefont {Landig}}, \bibinfo {author}
  {\bibfnamefont {G.}~\bibnamefont {Kucsko}}, \bibinfo {author} {\bibfnamefont
  {H.}~\bibnamefont {Zhou}}, \bibinfo {author} {\bibfnamefont {J.}~\bibnamefont
  {Isoya}}, \bibinfo {author} {\bibfnamefont {F.}~\bibnamefont {Jelezko}},
  \bibinfo {author} {\bibfnamefont {S.}~\bibnamefont {Onoda}}, \bibinfo
  {author} {\bibfnamefont {H.}~\bibnamefont {Sumiya}}, \bibinfo {author}
  {\bibfnamefont {V.}~\bibnamefont {Khemani}}, \bibinfo {author} {\bibfnamefont
  {C.}~\bibnamefont {{von Keyserlingk}}}, \bibinfo {author} {\bibfnamefont
  {N.~Y.}\ \bibnamefont {Yao}}, \bibinfo {author} {\bibfnamefont
  {E.}~\bibnamefont {Demler}}, \ and\ \bibinfo {author} {\bibfnamefont {M.~D.}\
  \bibnamefont {Lukin}},\ }\href {\doibase 10.1038/nature21426} {\bibfield
  {journal} {\bibinfo  {journal} {Nature}\ }\textbf {\bibinfo {volume} {543}},\
  \bibinfo {pages} {221} (\bibinfo {year} {2017})}\BibitemShut {NoStop}%
\bibitem [{\citenamefont {{Sherson}}\ \emph {et~al.}(2010)\citenamefont
  {{Sherson}}, \citenamefont {{Weitenberg}}, \citenamefont {{Endres}},
  \citenamefont {{Cheneau}}, \citenamefont {{Bloch}},\ and\ \citenamefont
  {{Kuhr}}}]{Sherson2010}%
  \BibitemOpen
  \bibfield  {author} {\bibinfo {author} {\bibfnamefont {J.~F.}\ \bibnamefont
  {{Sherson}}}, \bibinfo {author} {\bibfnamefont {C.}~\bibnamefont
  {{Weitenberg}}}, \bibinfo {author} {\bibfnamefont {M.}~\bibnamefont
  {{Endres}}}, \bibinfo {author} {\bibfnamefont {M.}~\bibnamefont {{Cheneau}}},
  \bibinfo {author} {\bibfnamefont {I.}~\bibnamefont {{Bloch}}}, \ and\
  \bibinfo {author} {\bibfnamefont {S.}~\bibnamefont {{Kuhr}}},\ }\href
  {\doibase 10.1038/nature09378} {\bibfield  {journal} {\bibinfo  {journal}
  {Nature}\ }\textbf {\bibinfo {volume} {467}},\ \bibinfo {pages} {68}
  (\bibinfo {year} {2010})}\BibitemShut {NoStop}%
\bibitem [{\citenamefont {{Beterov}}\ \emph {et~al.}(2009)\citenamefont
  {{Beterov}}, \citenamefont {{Ryabtsev}}, \citenamefont {{Tretyakov}},\ and\
  \citenamefont {{Entin}}}]{Beterov2009}%
  \BibitemOpen
  \bibfield  {author} {\bibinfo {author} {\bibfnamefont {I.~I.}\ \bibnamefont
  {{Beterov}}}, \bibinfo {author} {\bibfnamefont {I.~I.}\ \bibnamefont
  {{Ryabtsev}}}, \bibinfo {author} {\bibfnamefont {D.~B.}\ \bibnamefont
  {{Tretyakov}}}, \ and\ \bibinfo {author} {\bibfnamefont {V.~M.}\ \bibnamefont
  {{Entin}}},\ }\href {\doibase 10.1103/PhysRevA.79.052504} {\bibfield
  {journal} {\bibinfo  {journal} {Phys. Rev. A}\ }\textbf {\bibinfo {volume}
  {79}},\ \bibinfo {pages} {052504} (\bibinfo {year} {2009})}\BibitemShut
  {NoStop}%
\bibitem [{\citenamefont {Schachenmayer}\ \emph {et~al.}(2010)\citenamefont
  {Schachenmayer}, \citenamefont {Lesanovsky}, \citenamefont {Micheli},\ and\
  \citenamefont {Daley}}]{Schachenmayer2010}%
  \BibitemOpen
  \bibfield  {author} {\bibinfo {author} {\bibfnamefont {J.}~\bibnamefont
  {Schachenmayer}}, \bibinfo {author} {\bibfnamefont {I.}~\bibnamefont
  {Lesanovsky}}, \bibinfo {author} {\bibfnamefont {A.}~\bibnamefont {Micheli}},
  \ and\ \bibinfo {author} {\bibfnamefont {A.~J.}\ \bibnamefont {Daley}},\
  }\href {\doibase 10.1088/1367-2630/12/10/103044} {\bibfield  {journal}
  {\bibinfo  {journal} {New J. Phys.}\ }\textbf {\bibinfo {volume} {12}},\
  \bibinfo {pages} {103044} (\bibinfo {year} {2010})}\BibitemShut {NoStop}%
\end{thebibliography}%

\section*{Supplementary Information}
\section{Experimental details}

We started our experiment with a degenerate two-dimensional gas of rubidium-$87$ atoms, 
spin polarized in the hyperfine state $\ket{F,~m_F}=\ket{1,-1}$ and held in a single
confining antinode of an optical lattice along the $z$-direction.
We then switched on a square optical lattice with $\alat=532\,$nm spacing in
this single $x$-$y$-plane, preparing about $170$ atoms in a unity-filling Mott
insulator. For the experimental sequence, we ramped the optical lattices along
the $(x,y,z)$-directions to depths of $(40,40,80)E_r$, where
$E_r=\frac{h^2}{8m\alat^2}$ is the recoil energy for rubidium-$87$ in our
lattice. In this regime, the spatial extent of the Wannier function of
an atom in a lattice site as well as tunnelling between the sites on the
timescale of the experiment can be neglected and the Mott insulator served as a well controlled
starting state for our experiments with a single spin per site.
Employing an addressing scheme based on a digital mirror device
~\cite{Weitenberg2011a,Fukuhara2013}, we selected a single line of $10$
atoms aligned with the $x$-direction in the lattice for our experiment,
optically removing all other atoms from the trap. The filling of $87(3)\%$
of a single line is determined by infidelities in the single atom addressing
scheme and the filling of the initial Mott insulator ($90\%$ and $97\%$ respectively).
In order to introduce long-range interactions, the state $\ket{F=2,~m_F=-2}$ was
optically coupled to the $31P_{1/2}$, $\ket{J=1/2,~m_J=+1/2}$ Rydberg state.
The coupling laser at a wavelength of $298\,$nm was $\sigma_+$-polarized and 
propagated in the plane of the atoms at an angle of $45^{\circ}$
with the initially prepared chain of atoms.
We used approximately $77(10)\,$mW of uv-light, focused down to a
waist of $18(3)\,\mu$m to achieve a Rabi frequency of $\Omega/2\pi=3.57(3)\,$MHz
on the Rydberg transition. The Rabi frequency was calibrated by measuring the
AC-Stark shift $\delta=\frac{\Omega^2}{4\Delta}$ of the dressed ground state
$\ket{2,-2}$ due to the Rydberg dressing laser for different detunings
$\Delta$~\cite{Zeiher2016a}.
A magnetic field of strength $B_{xy}=0.405\,$G was applied in the plane of the atoms
along the excitation laser to set a well defined quantization axis for the
dressing scheme and the ground state spin basis.

\begin{figure}
  \centering
  \includegraphics{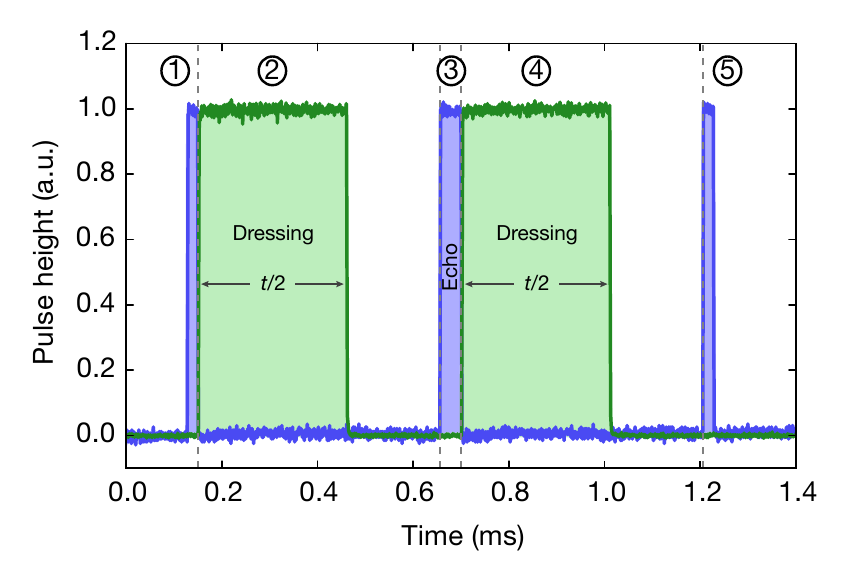}%
  \caption{\label{fig:5}
  \textbf{Experimental pulse sequence.}
  Microwave pulse power and dressing pulse power recorded using a photodiode, normalized to the maximum, are shown in blue and green. The role of the pulses in the experimental sequence was the preparation of all spins in an equal superposition of the states $\ket{1,-1}$ and $\ket{2,-2}$ in an eigenstate of $\hat{S}^y$ ($1$), followed by a first dressing interval of time $t/2$ ($2$). A spin echo pulse ($3$) was used to cancel phases acquired by every spin due to the dressing laser light shift after the second dressing interval of time $t/2$ ($4$). The final spin-readout along the $S^y$-direction was realized by a global spin rotation ($5$) identical to the one used for the preparation. To ensure the same decoherence due to magnetic field noise for measurements with different dressing times $t$, the time between the two $\pi/2$ pulses ($1,5$) was kept constant. 
The microwave pulses coupling $\ket{1,-1}$ and $\ket{2,-2}$ with an area $\pi/2$ and $\pi$ lasted $22\,\mu$s and $44\,\mu$s respectively. The dephasing time due to drifts of the global magnetic field was $2.8(4)\,$ms}
\end{figure}

\begin{figure*}
  \centering
  \includegraphics{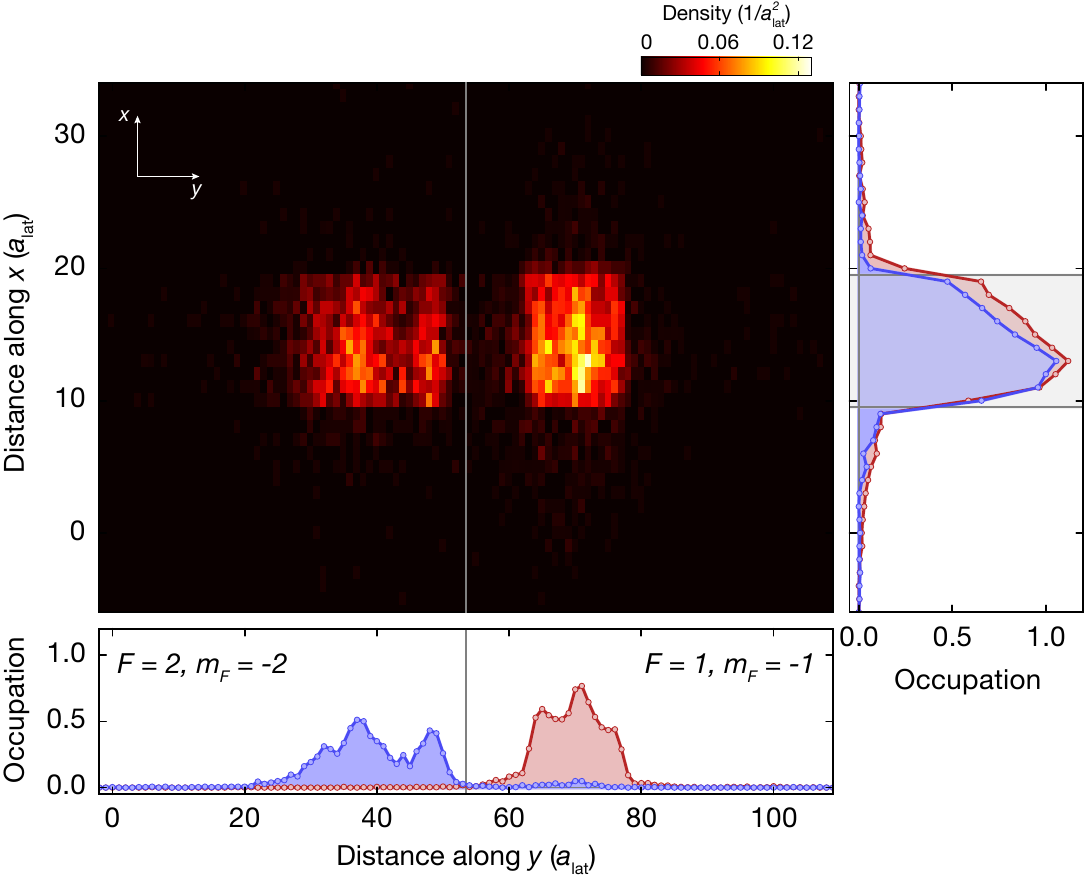}%
  \caption{\label{fig:6} 
  \textbf{Spin-resolved detection of spin chains.} 
  To detect the number of spins in a chain in the states $\ket{1,-1}$ and $\ket{2,-2}$, a magnetic field gradient was applied in the plane of the atoms along the $y$-direction, orthogonal to the initially prepared chain, to separate the two components due to their different magnetic moments. The main figure shows the density of atoms obtained after this procedure as an average over many experimental shots. The panel on the right shows the average of the density along the $y$-direction for atoms in $\ket{1,-1}$ and $\ket{2,-2}$ (red and blue, respectively). The gray shading marks the region of interest chosen for extracting the number of atoms in the two states. Its width of $10$ sites matches the width of the initially addressed chain. The lower panel equivalently shows the density averaged over the $x$-direction. The gray vertical line marks the cut to distinguish between atoms in $\ket{1,-1}$ (to the right) and $\ket{2,-2}$ (to the left). The position of the cut was obtained by minimizing the probability of false positive detection events for spin polarized initial states. For the chosen cut, we estimate the probability of an atom in $\ket{1,-1}$ to be falsely counted as a $\ket{2,-2}$ atom to be approximately $2\%$. The slight shift to the left of the central minimum is due to atoms in $\ket{1,-1}$ which were not perfectly removed in the preparation sequence of the single line. All results obtained in the main text are insensitive to small shifts of the separating line by $\pm\alat$.}
\end{figure*}

\subsection{Pulse sequence}
The experimental sequence for measuring the collapse and revival dynamics is shown in Fig.~\ref{fig:5}. The cancellation of the dressing laser induced AC-Stark shift was of paramount importance for the successful detection of collapse and revivals in our experiment because it can, if not controlled, lead to a fast dephasing of the spin dynamics. This was accomplished by sandwiching the two dressing phases of duration $t/2$ in a microwave echo sequence, where the single particle shifts acquired during the two dressing intervals before and after the spin-echo pulse are cancelled, provided the two dressing pulse areas are equal. The total time between the two microwave pulses of area $\pi/2$ was kept constant to ensure the same effect of magnetic field decoherence for all dressing times $t$. 
For our parameters $\Omega/2\pi=3.57(1)\,$MHz, $\Delta/2\pi=11.00(2)\,$MHz, the dressing laser induced AC-Stark shift amounted to $\delta/2\pi\approx290\,$kHz, exceeding the nearest-neighbor interaction $|U_0|=13.1(5)\,$kHz by a factor of $22$. Hence, for the revival at $t=5/U_0$, a small fluctuation of the dressing pulse areas between the two intervals of $1\%$ already leads to a detrimental phase error of more than $\pi$ in the global spin. The detection of the revival in our experiments shows that the two areas of the dressing laser pulses shown in Fig.~\ref{fig:5} cancelled to a better degree in our experiment.
Nevertheless, to eliminate experimental runs with excessive pulse fluctuations, we monitored the dressing pulses for each experimental run and only used those shots for the analysis, where the relative difference of the two pulse areas was smaller than $1\%$ and the absolute height of the pulse deviated from the mean of all pulse heights by less than $2.5\%$. This eliminated $5\%$ and $7\%$ of the dataset, respectively.
In addition, we used the recorded total pulse area to rescale the dressing time for a run $k$ by the relation $t_k=(h_k/\bar{h})^2\tilde{t}$, where $h_k$ is the intensity measured by the photo diode for shot $k$, $\bar{h}$ is the mean pulse height averaged over all experimental shots ($3880$ in total) and $\tilde{t}$ denotes the set length of the pulses.  

\subsection{Spin-resolved detection}
\begin{figure}
  \centering
  \includegraphics{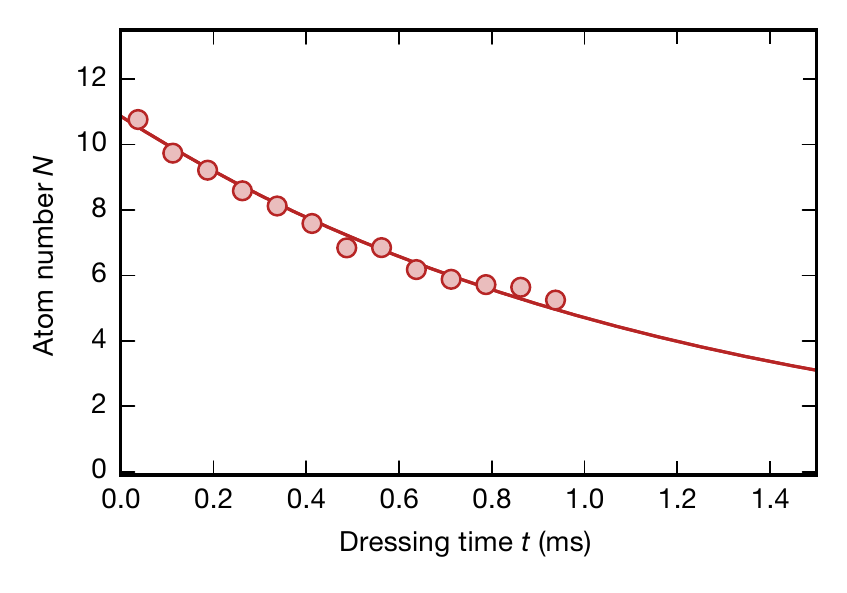}%
  \caption{\label{fig:7} 
  \textbf{Lifetime of the dressed spin chain.} 
  Measured atom number after variable dressing time $t$ in the echo sequence. The lifetime extracted from an exponential fit is $\tau=1.20(4) \,$ms. Error bars denote the s.e.m. and are smaller than the size of the symbols.}
\end{figure}

After the second dressing interval, we applied a microwave $\pi/2$-pulse to rotate the spins to the measurement basis in the $S^z$-direction. Here, the populations of the two eigenstates $\ket{1,-1}$ and $\ket{2,-2}$ can be detected due to their different magnetic moments. To achieve the latter, we allowed tunnelling orthogonal to the initially prepared chain by ramping down the $y$-lattice. At the same time, we adiabatically ramped up a magnetic field gradient in the plane of the atoms. This led to a separation of the two spin states along the $y$-direction, see Fig.~\ref{fig:6}. After the atoms had settled to their respective new equilibrium position, we ramped up the $y$-lattice, followed by a fluorescence image to obtain the site-resolved occupation of each lattice site~\cite{Sherson2010}.
The splitting procedure allowed the detection of the spin state of an atom with approximately $98(1)\%$ and was mainly limited due to residual atoms in the initial two-dimensional Mott insulator which had not been removed in the addressing scheme and whose distribution had a small overlap with the $\ket{2,-2}$ spins after the splitting, see Fig.~\ref{fig:6}.

\begin{figure*}
  \centering
  \includegraphics{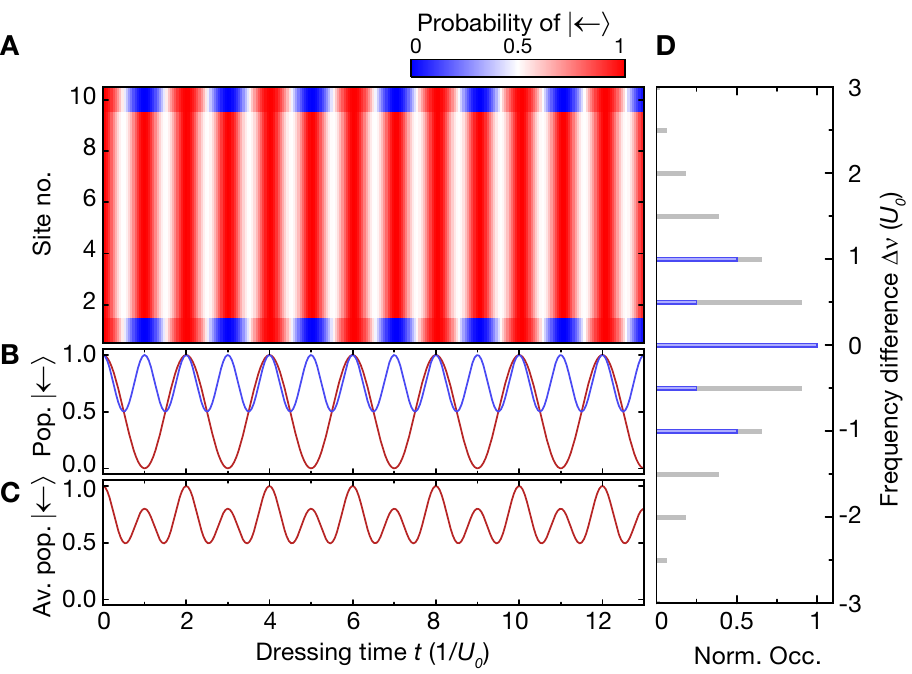}%
  \caption{ \label{fig:8} 
  \textbf{Simulation of spatially resolved magnetization dynamics for nearest-neighbor Ising interactions.}
  \textbf{(A)} Theoretically calculated evolution of the probability (color coded) for a spin at a given lattice site (labelled from $1$ to $10$) to point in the initially prepared $S^y$-direction along $\ket{\leftarrow}$ for each spin versus dressing time $t$ in a defect-free chain of $N=10$ atoms. The out-of-phase time evolution of the edge spin is clearly visible. All bulk spins display periodic revivals at times $t=n/U_0$. \textbf{(B)} shows representative traces as cuts of (A) for a spin at the edge of the sample (red) and a spin in the bulk (blue), illustrating the different dynamics. \textbf{(C)} Probability for a spin to point in the initial $S^y$-direction along $\ket{\leftarrow}$ averaged over the full chain. Due to the different edge spin dynamics, every second revival is suppressed. \textbf{(D)} The frequency differences $\Delta\nu$ of the many-body eigenstates of the model are multiples of the nearest-neighbor interaction strength $U_0=-13.1(5)\,$kHz (gray), but only few are relevant for the dynamics of the transverse magnetization density (blue bars).}
\end{figure*}

\section{Lifetime and coherence of the Rydberg-dressed spin chain}
Due to the admixture of the Rydberg state with lifetime $\tau_R$ to the ground state $\ket{2,-2}$, the latter acquires a finite effective lifetime $1/\gamma=\tau_R/\beta^2$, where $\beta=\Omega/2\Delta$ denotes the admixed amplitude of the Rydberg state to the ground state. For our parameters $\beta^2=2.63(5)\%$ and the literature value $\tau_R=27\,\mu$s~\cite{Beterov2009}, we expect $1/\gamma=1.04(2)\,$ms under ideal conditions.
The decoherence induced by coupling to the Rydberg state results predominantly in loss from the trap to an unobserved state with rate $\gamma_0\approx\gamma$~\cite{Zeiher2016a}.
Experimentally, we directly obtain access to $\gamma_0$ in our system by measuring the time evolution of the total atom number $N$ in the experimental sequence used to study the magnetization dynamics. It is expected to decay exponentially by~\cite{Zeiher2016a} 
\begin{equation}
N(t)= N_0~e^{-\gamma_0t/2}\equiv N_0~e^{-t/\tau}.
\end{equation}
The factor of one half appearing in the exponent is a consequence of a reduced $\tilde{\beta}=\beta/\sqrt{2}$ for the states in the $S^y$-$S^x$ (equator) plane of the Bloch sphere. There, each state is an equal superposition of dressed and undressed ground state, only differing by a relative phase.
The measured atom number decay is shown in Fig.~\ref{fig:7}. An exponential fit to the data captures the decay well and yields a time constant of $\tau=1.20(4)\,$ms, reaching $60\%$ of the theoretically expected value $2/\gamma_0$ under ideal conditions.

The time evolution of the expectation value of the parity operator, $\langle\hat{P}\rangle = \langle e^{-i\pi\sum
\limits_{i=1}^{N}{\hat{S}^{\rightarrow}_i}} \rangle$, sheds additional light on decoherence processes present in the system. For short times $t/\tau\ll1$ where the atom number is close to its initial value $N_0$ and where we experimentally probed the parity, it can be shown following the formalism used in~\cite{Foss-Feig2013, Zeiher2016a} that $\langle\hat{P}\rangle$ decays as
\begin{equation}
\langle \hat{P}(t)\rangle= P_0~e^{-\frac{N}{2}(\gamma_0+\gamma_{\downarrow}+\gamma_{\uparrow})t}\equiv P_0~e^{-t/\tau_P}.
\end{equation}
In addition to decoherence by the above discussed atom loss, also spin flips from the dressed ground state $\ket{2,-2}$ to the undressed ground state $\ket{1,-1}$ with rate $\gamma_{\downarrow}$, and dephasing of the dressed state with rate $\gamma_{\uparrow}$ contribute to the decay of the parity.
Interestingly, the number of atoms $N$ in the system appears as a scaling factor in the exponent, making the parity a very sensitive probe for all three decoherence processes.
From the measurement (see Fig.~\ref{fig:4}) we extract $\tau_P/\tau\approx 1/N$, from which we conclude that $\gamma_{\downarrow}$ and $\gamma_{\uparrow}$ only contribute insignificantly to the total decoherence of the system.
This is also consistent with our observation that the magnetization density shows revival dynamics even when a fraction of the atoms has been lost, which indicates that no excessive dephasing is present in the system. 
The achieved long atom number lifetime $\tau$ in our experiment shows that collective decay effects limiting the achievable coherence times in two dimensions~\cite{Zeiher2016a} seem to be strongly suppressed in the 1d system. We experimentally checked the decay time $\tau$ for different detunings and verified the absence of collective losses for detunings as small as $\Delta/2\pi=4\,$MHz. This promises far longer coherence times when working closer detuned to resonance, as the interaction-lifetime product increases and the influence of first order AC-Stark shifts decreases relative to the interactions strength with decreasing detuning.

\begin{figure*}
  \centering
  \includegraphics{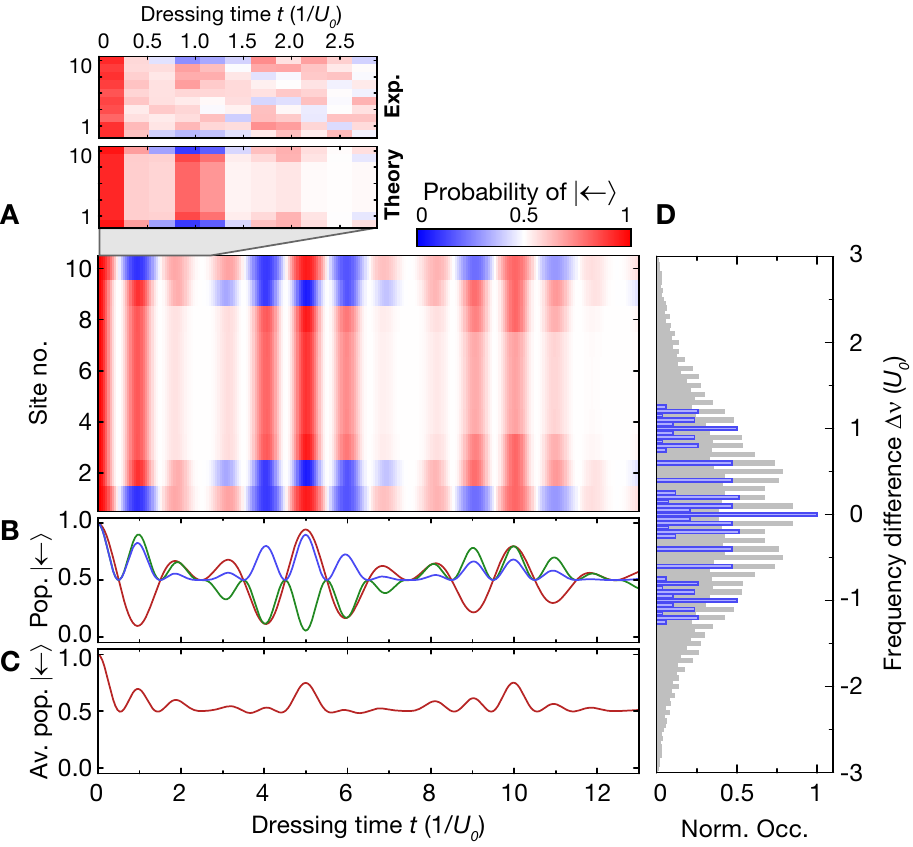}%
  \caption{ \label{fig:9} 
  \textbf{Spatially resolved magnetization dynamics for Rydberg-dressed interactions.}
  \textbf{(A)}  Theoretically calculated evolution of the probability (color coded) for a spin at a given lattice site (labelled from $1$ to $10$) to point in the initially prepared $S^y$-direction along $\ket{\leftarrow}$ for each spin versus dressing time $t$ in a defect-free chain of $N=10$ atoms. Contrary to the nearest-neighbor interacting Ising chain, the bulk spin-dynamics is modulated by the next-nearest-neighbor interaction $U(2\alat)\approx U_0/5$. Near the edge, spins show different temporal dynamics and now also the spins further in the chain are affected by the edge. Above, a zoom  into the initial part of the dynamics, indicated by the gray area, shows the measured site-resolved probability for a spin at a given lattice site to point in the initially prepared $S^y$-direction along $\ket{\leftarrow}$ versus dressing time $t$ (upper panel) and the corresponding theoretical expectation for a defect-free initial chain using the same binning as for the experimental data (lower panel).
  \textbf{(B)} illustrates the dynamics with single spin time traces for an edge spin (red), the spin neighboring the edge spin (green) and spins in the bulk (blue), all taken as cuts of (A). \textbf{(C)} The average global spin now displays an interaction-induced collapse after two weak, bulk driven revivals at $t=1/U_0$ and $t=2/U_0$, before reviving at $t=5/U_0$ and later at $t=10/U_0$. \textbf{(D)} The frequency differences $\Delta\nu$ of the many-body eigenstates of the long-range interacting chain (gray bars) and those relevant for the dynamics of the transverse magnetization density (blue bars). Compared to the nearest-neighbor interacting Ising chain, many more frequencies contribute.} 
\end{figure*}

\section{Dynamics including the spin echo}
In this section, a link between the time evolution including a spin echo and the evolution in a rotating frame of reference is established. To this end, we first consider the exact model realized in the experiment, where only spins in the optically dressed state $\ket{2,-2}$ interact. It is given as
\begin{equation}
\label{eq:S1}
\hat{H_a}=h\sum_{i\neq j}^{N}\frac{U(d_{ij})}{2}~\hat{\sigma}^e_i \hat{\sigma}^e_j.
\end{equation}
Here, $\hat{\sigma}^e_i = \ket{e}\bra{e}$ measures the occupation of the dressed state $\ket{e}=\ket{2,-2}$ and $U(d_{ij})$ with $d=|i-j|$ denotes the dressed interaction potential between two dressed atoms in state $\ket{e}$ at sites $i$ and $j$ respectively.   
This atomic model can be rewritten in terms of an Ising model~\cite{Schachenmayer2010,Schauss2015}, by introducing the spin operators $\hat{S}^z_i=\frac{1}{2}(2\hat{\sigma}^e_i-\mathbb{I})$, yielding
\begin{equation}
\label{eq:S2}
\hat{H}_{\mathrm{rot}}=h\sum_{i\neq j}^{N}\frac{U(d_{ij})}{2}~\hat{S}^z_i \hat{S}^z_j + h\sum_{i}^{N} \Delta^{\mathrm{(coll)}}_i\hat{S}^z_i \equiv \hat{H} + \hat{H}_s.
\end{equation}
The term $\Delta^{\mathrm{(coll)}}_i=\sum_{i\neq j}^{N} \frac{U(d_{ij})}{2}$ is picked up additionally in this transformation and acts as an additional longitudinal magnetic field, captured by the Hamiltonian $\hat{H}_s$. 
It is linear in the spin operator $\hat{S}^z_i$ and could hence be eliminated in a rotating frame of reference generated by the unitary operator $\hat{U}=e^{-i 2\pi t\sum_{i}^{N} \Delta^{\mathrm{(coll)}}_i\hat{S}^z_i}=\bigotimes_i^{N} e^{-i2\pi t\Delta^{\mathrm{(coll)}}_i\hat{S}^z_i} = \bigotimes_i^{N} \hat{U}_i$. However, due to the spatial dependence on $i$ in a finite system, local control of the rotation operators $\hat{U}_i$ would be required. 
An alternative, experimentally more tractable strategy for finite systems is the introduction of spin echo pulses of the form $\Pi=\bigotimes_j^N e^{-i\pi\hat{S}^x_j}$. These spin echoes can be implemented by a global microwave pulse, applied after half of the evolution time $t$. They effectively invert the roles of $\ket{\uparrow}$ and $\ket{\downarrow}$ and hence eliminate the effects of terms linear in $\hat{S}^z_i$ during the time evolution, leaving the part $\hat{H}$ of the Hamiltonian with the spin-spin interactions as the only drive for the dynamics.
In order to prove that the time evolutions of $\hat{H}_{\mathrm{rot}}$ and $\hat{H}$ are equivalent if a spin echo is applied after an evolution time $t/2$, we evaluate
\begin{align*}
e^{-i t\hat{H}_{\mathrm{rot}}/2\hbar}~\Pi~e^{-i t\hat{H}_{\mathrm{rot}}/2\hbar} 
&= e^{-i t(\hat{H}+\hat{H}_s)/2\hbar}~\Pi~e^{-i t(\hat{H}+\hat{H}_s)/2\hbar}\\
&=\Pi~e^{-i t(\hat{H}-\hat{H}_s)/2\hbar} e^{-i t(\hat{H}+\hat{H}_s)/2\hbar}\\
&=\Pi~e^{-i t\hat{H}/\hbar}.
\end{align*}
In the second step we have used the anti-commutation relation for the spin operators at the same site $i$, $\{\hat{S}^z_i,\hat{S}^x_i\}=0$.
This shows that the time evolution of the system under $\hat{H}_{\mathrm{rot}}$ is equivalent to the one under $\hat{H}$, up to the spin echo pulse, which merely amounts to a global phase and a redefinition of the measurement basis and has no influence on the dynamics.
 
\section{Spatially resolved dynamics of the revivals}
For the system studied in our experiment, the expected spin dynamics can be calculated exactly. This allows to obtain both expectation values for mean spin, spin-spin correlations and the spatial structure of the revivals~\cite{Foss-Feig2013,Richerme2014,Zeiher2016a}. As mentioned in the main text, the expectation value for a spin at a site $j$ can be evaluated to yield 
\begin{equation}
\label{eq:S3}
\langle\hat{S}^y_j(t)\rangle=\frac{1}{2}\prod\limits_{i\neq j}^{N}\cos(\pi U(d_{ij})t).
\end{equation}
The dynamics of the spin is governed by the interaction with its neighbors and the resulting beat notes due to frequency mixing. In the following, we will link this analytic result for the local magnetization dynamics to the spectral features of the Hamiltonian 
\begin{equation}
\hat{H}=h\sum_{i\neq j}^{N}\frac{U(d_{ij})}{2}~\hat{S}^z_i \hat{S}^z_j. 
\end{equation}
To first obtain an intuitive understanding of the dynamics, we consider the time evolution of an Ising model with an interaction potential $U(d_{ij})=U_0$ constrained to nearest-neighbor interaction only. In this case, the mean magnetization evolves with periodic revivals at times $t=n/U_0$, multiples of the interaction time $1/U_0$ (see Fig.~\ref{fig:8}C). 
The frequency differences dominating the dynamics are limited to $\Delta\nu = 0, \pm U_0, \pm U_0/2$. This can be understood from the simple structure of the many-body spectrum of $\hat{H}$, whose set of eigenstates $\ket{\lambda}$ comprises all products of the single spin eigenstates $\ket{\uparrow}$ and $\ket{\downarrow}$ of $\hat{S}^z$.
The spectrum of $\hat{H}$ only allows for $\Delta\nu$ being a multiple of $\pm nU_0/2$ (Fig.~\ref{fig:8}D). The matrix element $\expect{\eta}{\hat{S}^y_j}{\lambda}$ is non-zero for those many-body eigenstates $\ket{\eta}$ and $\ket{\lambda}$ which differ by a single flipped spin from $\ket{\uparrow}$ to $\ket{\downarrow}$ at the same position $j$. The cost of such a flipped spin is measured by $\Delta\nu$ and amounts to $\Delta\nu=0$ for anti-aligned neighbors of site $j$ ($\ket{\dots\downarrow\uparrow\uparrow  \dots}\longrightarrow\ket{\dots\downarrow\downarrow\uparrow  \dots}$, the central spin is assumed to be at site $j$) or $\Delta\nu=\pm U_0$ if neighbors are aligned ($\ket{\dots\uparrow\uparrow\uparrow  \dots}\longrightarrow\ket{\dots\uparrow\downarrow\uparrow  \dots}$ or $ \ket{\dots\downarrow\downarrow\downarrow  \dots} \longrightarrow\ket{\dots\downarrow\uparrow\downarrow  \dots}$).  
From this argument, it is also clear that the magnetization of the edge spin will evolve differently due to its different environment. Focusing on the spatial structure of the magnetization dynamics shown in Fig.~\ref{fig:8}A, this is directly visible. Due to the single missing neighbor, the spin flip energy $\Delta\nu$ is only $\pm U_0/2$ and hence the oscillation is correspondingly slower than that of a bulk spin. As a consequence, the strength of every second revival of the average total magnetization of the chain is reduced, since the edge spins are out-of-phase.

Similar arguments hold for the long-range interaction potential. However, here the interaction spectrum is much more complicated due to beyond nearest-neighbor interactions and the number of relevant frequency differences $\Delta\nu$ increases significantly, acquiring sidebands due to interactions with spins at larger distances (see Fig.~\ref{fig:9}D). Therefore, the revival dynamics is more complex. First, the fast oscillations of the bulk with periodicity $1/U_0$, showing full contrast in the nearest-neighbor Ising case, are modulated with the next-nearest-neighbor interaction. This leads to damping of the magnetization revival amplitude initially, but allows for the partial revivals later at $t=5/U_0$ and $t=10/U_0$. Second, the edge spin and the spin next to the edge spin show different dynamics compared with the bulk, see Fig.~\ref{fig:9}B, which can again be understood from the frequencies contributing to $\langle\hat{S}^y_j\rangle$. The initial out-of phase-dynamics of the edge spin is also observed experimentally in the spatially resolved magnetization density, see Fig.~\ref{fig:9}A. Summing the time evolution traces for the mean local magnetization $\langle\hat{S}^y_j\rangle$ yields the total collapse and revival dynamics with the characteristic features observed also in the experiment (see Fig.~\ref{fig:9}C).
As a final remark it should to be noted that the different evolution of the edge spins is has similar dynamical features as the ``collective field'' $\Delta^{\mathrm{(coll)}}_i$~\cite{Zeiher2016a} resulting from the symmetrization of the spin-spin interaction potential. However, contrary to the collective field it is not a mere single particle effect and it is therefore not removed by the spin-echo pulse, but rather a consequence of the direct spin-spin interaction and the finite system size.

\end{document}